\documentclass[10pt,a4paper]{article}

\usepackage[T1]{fontenc}
\usepackage[utf8]{inputenc}
\usepackage{amssymb,amsmath,amsthm,amsfonts}    
\usepackage{bm}                                 
\usepackage{mathtools}                          
\usepackage{stmaryrd}                           
\usepackage{breqn}                              
\usepackage[textfont=footnotesize,labelfont={footnotesize,bf}]{caption}
\usepackage{subcaption}
\usepackage{booktabs}
\usepackage[inline]{enumitem}
\usepackage{float}
\usepackage{graphicx}
\usepackage{graphbox}
\usepackage[dvipsnames]{xcolor}
\usepackage[top=30mm,right=30mm,left=30mm,bottom=30mm]{geometry}
\usepackage[colorlinks=true,linkcolor=blue,urlcolor=blue]{hyperref}
\usepackage[affil-it,auth-sc]{authblk}
\usepackage[colorinlistoftodos,textwidth=25mm,textsize=footnotesize]{todonotes}
\usepackage{lipsum}                             
\usepackage[section]{placeins}
\usepackage{tikz}
\usetikzlibrary{positioning}
\usepackage{scalefnt}
\usepackage[
	backend=biber,
	style=numeric,
	citestyle=numeric-comp,
	maxbibnames=99,
	minbibnames=99,
	maxcitenames=2,
	mincitenames=1,
	giveninits=true,
	sorting=none,
	sortlocale=auto,
	natbib=true,
	url=false,
	isbn=false,
	doi=true,
	eprint=true
]{biblatex}
\graphicspath{{./}{./figs}}                 

\title{Three Homogenization Models for a Nonlinear Heterogeneous Rod with Microscale Dead Loading}

\author[1]{Yang Ye}
\author[1]{Andrea Piccolroaz} 
\author[1]{Davide Bigoni\footnote{Corresponding author: e-mail: \href{mailto:bigoni@unitn.it}{bigoni@ing.unitn.it}; phone: +39\,0461\,282507.}}

\affil[1]{Instabilities Lab, University of Trento, Italy}

\date{}

\begin{document}

\maketitle

\begin{abstract}
A discrete chain of one-dimensional elements with concentrated elastic stiffness is analyzed to investigate the effect of self-equilibrated dead loads acting at the microscale. Three homogenization schemes are formulated for a finite chain whose total length is much larger than the length of an individual cell.

Model (i) homogenizes the elastic properties and the microscopic loading simultaneously, yielding an equivalent rod with homogeneous axial and flexural stiffnesses. The effect of the microforces is incorporated through an effective prestress state and a distribution of nonlinear rotational springs accounting for their smeared contribution. Two additional models are developed in which the microforces are not homogenized but remain periodically distributed: Model (ii), an equivalent rod with homogeneous axial and flexural stiffnesses, and Model (iii), a flexurally homogeneous equivalent rod with piecewise-constant axial stiffness.

Although Model (iii) most closely resembles the original discrete mechanical system and Model (ii) may appear the most intuitive from an engineering standpoint, Model (i) is shown to recover the continuum limit of the fully loaded discrete system and to provide the most accurate, fully homogenized description. Moreover, the explicit incorporation of the microforces into the nonlinear homogenization procedure of Model (i) gives rise to a form of loading and distributed constraint in a continuous rod that has no direct counterpart in conventional mechanical systems. In particular, it can produce an effective distribution of rotational springs with negative stiffness.

The nonlinear equilibrium equation of Model (i) also admits a closed-form solution in terms of elliptic functions, which is used to analyze the large-deflection response and the initial post-buckling behavior of a cantilever. The model predicts both tensile and compressive buckling, as well as restabilization of the straight configuration under suitable loading conditions.

Finally, the combined control of the end load and the microforces is explored as a mechanism for soft-robotic actuation. Suitable loading paths allow the system to switch between straight and bent configurations and to modify the bifurcation character from supercritical to subcritical.
\end{abstract}

\paragraph{Keywords}
Metamaterials \textperiodcentered\ 
Homogenization \textperiodcentered\
Prestress \textperiodcentered\
Material instabilities

\section{Introduction}

The emergence of advanced automated manufacturing has opened the way to materials and structures engineered with increasingly sophisticated microstructural designs.
Such designs enable the realization of microstructured chains of elements with lumped elastic stiffness that, once homogenized, can be described as equivalent continuous rods in which axial, shear, and flexural stiffnesses are independent and may, in principle, assume arbitrary values.
This capability represents a new paradigm and an opportunity for the development of one-dimensional rod elements with unprecedented mechanical properties \cite{frenzel_three-dimensional_2017, frenzel_large_2021, iglesias_martinez_experimental_2021}. 

Complementary to the possibilities outlined above, Challamel and co-workers \cite{Kocsis, kocsis_foundation_2018} developed a nonlinear homogenization technique for deriving equivalent continuous rods from periodic chains of finite length, rather than from infinite periodic media, building on earlier results by Domokos \cite{Domokos1993, Domokos1993109}. They showed that the Engesser and Haringx--Reissner shearable rod models arise as the homogenized response of different microstructural elements; this framework was recently extended in \cite{paradiso_nonlinear_2025}.

In the present article, a new microstructural chain is introduced to investigate the independent control of a macroscopic end force and a periodic distribution of microforces applied to each element. 
The microforces are dead loads acting axially, with zero resultant in the straight configuration taken as the reference. The crucial point is how these microforces enter the homogenization scheme. In fact, the following three different homogenization schemes are developed and shown to yield different results.
\begin{itemize}   
\item[(i)] The discrete chain and the microforces are homogenized simultaneously, yielding an equivalent rod that is homogeneous in both stiffness and internal forces. In particular, rather than appearing as externally applied loads, the microforces contribute to the internal prestress state of the rod and additionally give rise to a distribution of nonlinear rotational springs along its axis.
\item[(ii)] The discrete chain is homogenized into an equivalent rod capturing its axial and flexural deformability, while the microforces are neglected at this stage and later applied directly to the rod as external loads. The equivalent rod is elastically homogeneous and subject to a piecewise constant and periodic axial force in its straight configuration.
\item[(iii)] Flexural behavior is homogenized continuously in the equivalent rod, whereas the axial stiffness remains piecewise constant, periodically alternating between inextensible and axially deformable segments. The microforces are then applied to the equivalent rod as external loads, precisely at the junctions between the inextensible and deformable segments. 
\end{itemize}
The performance of the different homogenization schemes is assessed against the discrete structure, which is taken as the reference. Model (ii) is probably the simplest and most natural choice for homogenization. Relative to the discrete structure, where axially rigid and axially deformable segments alternate, Model (iii) constitutes a refinement of Model (ii). Both models lead to nonlinear governing equations whose internal forces are piecewise smooth and exhibit finite jumps; their analysis is feasible but technically 
involved\footnote{
To illustrate some of the technical difficulties associated with solving the piecewise equations, the buckling loads of a clamped rod defined according to Models (ii) and (iii) are both determined using a transfer matrix technique.
}
.

The comparison among the three models shows that the most straightforward formulation, Model (ii), provides the least satisfactory performance. In contrast, incorporating the microforces into the homogenization scheme, as in Model (i), combines relative simplicity with improved accuracy. 

Interestingly, Model (i) leads to a reinterpretation of the microforces: rather than acting as externally applied loads, they enter the homogenized model as a combination of an internal prestress determined by the boundary conditions and a distribution of rotational springs acting on the rod even when it is unstretched. 
Consequently, within the homogenized description, the microforces no longer act as external loads. Instead, they give rise to two unconventional continuum features: a novel prestress state and a distributed rotational constraint, neither of which can be realized through conventional loading mechanisms. The effective load governing the prestress combines two distinct contributions, whose relative magnitudes can be tuned to induce different responses, including instability and subsequent restabilization. Moreover, the stiffness of the distributed rotational springs can be tuned to attain negative values, enabling mechanical effects unavailable in conventionally loaded rods.

A crucial advantage of Model (i) is that it also permits the use of analytical techniques that are available only for rods free of distributed loads, unlike Models (ii) and (iii). In particular, Model (i) allows closed-form solutions to be derived for the nonlinear large-deflection response of the rod under end loading. Moreover, for the buckling problem, a weakly nonlinear analysis can be carried out to assess the character of the initial post-critical response. These analytical treatments are not available for the discrete chain, nor for the equivalent rod formulations (ii) and (iii).

Finally, the micromechanical model and its best-performing equivalent rod model are examined to assess their ability to generate unconventional mechanical responses and to explore their potential application as a soft robotic arm \cite{armanini_elastica_2017, armanini_soft_2023,sipos_longest_2020,falope_finite_2022, dhanakoti_stability_2025}. Although such applications would generally involve discrete chains comprising a large number of elements, all examples considered here deliberately employ only a small number of elements. This choice provides a particularly demanding test of the proposed homogenization technique, whose performance was consistently found to be remarkably accurate even under these severe conditions.

For a cantilever configuration, a complete bifurcation analysis is performed. The governing equations of the discrete model are integrated numerically to characterize the post-critical response, whereas an exact analytical solution is obtained for the equivalent rod model. The results reveal the possibility of both tensile \cite{zaccaria_tensile_2011} and compressive buckling, and in the latter case, the restabilization \cite{koutsogiannakis_restabilization_2023} of the straight configuration.

When interpreted as a soft robotic arm, the structure can undergo flexural cycles that start from the unloaded straight configuration and return to a straight configuration in a loaded, prestressed state. This behavior makes the structure particularly attractive for soft actuation applications.

\section{One-dimensional periodic structure loaded with dead microforces}
\label{sec:1D-chain-discrete}

\begin{figure}[H]
    \centering
    \includegraphics[width=0.75\linewidth]{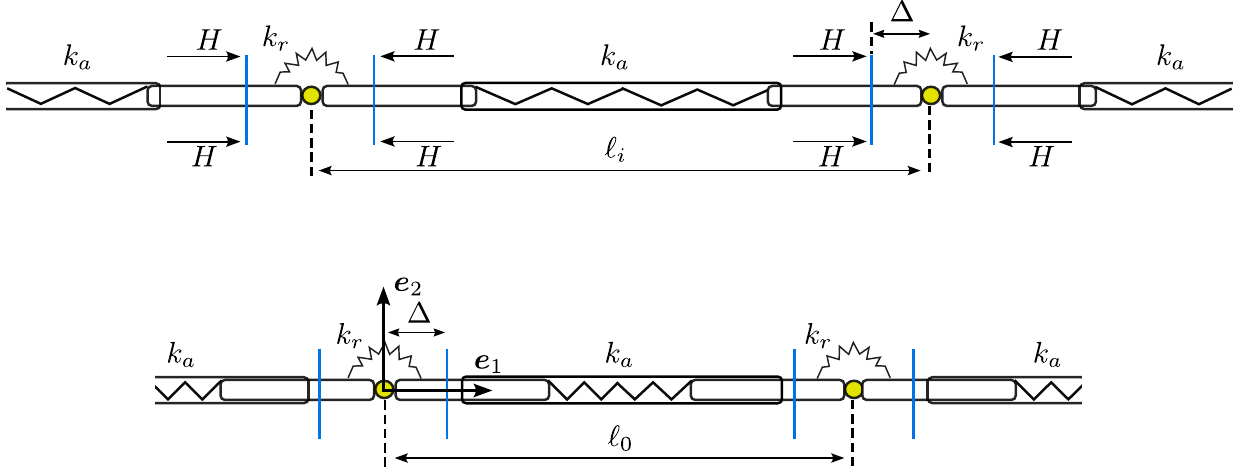}
    \caption{
    One-dimensional chain with lumped elasticity. In the lower figure, the chain is shown in its undeformed configuration. Flexural compliance is provided by hinges of rotational stiffness $k_r$, whereas axial deformation is confined to the central segment of axial stiffness $k_a$. A pair of equal and opposite dead forces, each of magnitude $2H$, acts within each element at the ends of the rigid segments of length $\Delta$, producing the same elongation in every unit cell, as shown in the upper figure.
    }
    \label{fig:1D-sketch}
\end{figure}

A one-dimensional discrete chain is considered, composed of axially deformable elements, each with axial stiffness $k_a$ and initial length $\ell_0$, that are flexurally rigid. The elements are connected by elastic hinges of rotational stiffness $k_r$, as shown in Fig.~\ref{fig:1D-sketch}. A pair of equal and opposite dead forces, each with a magnitude of $2H$, is applied within each element at a distance $\Delta$ from the hinges. For graphical clarity, these forces are shown as acting on the rigid bars marked in blue. The central portion, with an initial length $\ell_0-2\Delta$, is axially deformable, whereas the two end portions, each of length $\Delta$, are inextensible. 

The practical realization of the prescribed microforces lies beyond the scope of the present study, which focuses instead on their mechanical consequences within the proposed theoretical framework. Nevertheless, a possible strategy for generating dead forces of opposite sign has recently been presented in \cite{Bigoni2026106590}, providing a potential basis for the experimental implementation of such loading conditions.

\subsection{Kinematics of the unit cell}
\label{ssec:1D-chain-discrete-geometry-kinematics}

\begin{figure}[H]
    \centering
    \includegraphics[width=0.7\linewidth]{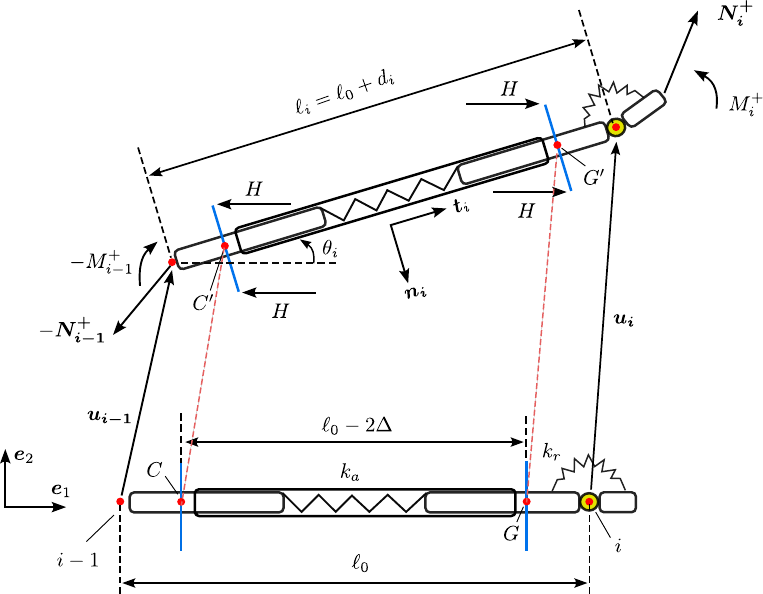}
    \caption{
    The $i$--th element in its undeformed configuration, with length $\ell_0$, and in its deformed configuration, with length $\ell_i=\ell_0+d_i$ and rotation $\theta_i$ relative to the unit vector $\bm{e}_1$. The forces and moments transmitted by the adjacent cells at the left end, $\bm{N}_{i-1}^+$ and $M_{i-1}^+$, and at the right end, $\bm{N}_i^+$ and $M_i^+$, are also shown. The microforces $2H$ remain horizontal because they are dead loads.
    }
    \label{fig:1D-horizontal-force-deformed}
\end{figure}

The microstructure shown in Fig.~\ref{fig:1D-sketch}, in its undeformed (lower part) and axially loaded (upper part) configurations, undergoes large deformations, as illustrated in Fig.~\ref{fig:1D-horizontal-force-deformed}. The figure shows a single unit cell subjected to the external forces $2H$ and the end forces transmitted by the adjacent cells. Denoting the displacement of the $i$--th node by $\bm{u}_i = u_i \bm{e}_1 + v_i \bm{e}_2$, where $\bm{e}_1$ and $\bm{e}_2$ are the horizontal and vertical unit vectors, respectively, the relative displacement of the nodes at the ends of the element is
\begin{equation}
\bm{u}_i - \bm{u}_{i-1} = (\ell_0 + d_i)\bm{t}_i - \ell_0\bm{e}_1, 
\label{eq:ui-slaved}
\end{equation}
where $d_i$ is the axial elongation of the element and $\bm{t}_i$ is the axial unit vector. The axial unit vector $\bm{t}_i$ and the unit normal $\bm{n}_i$ are defined as
\begin{equation}
\bm{t}_i = \cos\theta_i\, \bm{e}_1 + \sin\theta_i\, \bm{e}_2, \quad 
\bm{n}_i = \sin\theta_i\, \bm{e}_1 - \cos\theta_i\, \bm{e}_2, 
\label{eq:unit-tangent-normal}
\end{equation}
where $\theta_i$ is the rotation angle of the element. 

The current configuration of the $i$-th unit cell is described by four scalar kinematic variables: the two components of the displacement $\bm{u}_{i-1}$ of its left end, the rotation $\theta_i$, and the axial strain $\varepsilon_i = d_i/\ell_0$. The axial stretch $\lambda_i$ of the element is defined as
\begin{equation}
\lambda_i = 1 + \varepsilon_i = 1 + d_i/\ell_0. 
\label{eq:stretch-i}
\end{equation}
From Eq.~\eqref{eq:ui-slaved}, the following compatibility equations can be derived
\begin{equation}
\begin{aligned}
    \lambda_i &= \left(1 + \frac{u_i - u_{i-1}}{\ell_0}\right) \cos\theta_i + \frac{v_i - v_{i-1}}{\ell_0} \sin\theta_i, \\[2mm]
    \gamma_i &= -\left(1 + \frac{u_i - u_{i-1}}{\ell_0}\right) \sin\theta_i + \frac{v_i - v_{i-1}}{\ell_0} \cos\theta_i = 0, 
\end{aligned}
\end{equation}
where the first relation expresses the axial stretch in terms of the nodal displacements and the element rotation, whereas the second enforces the \emph{unshearability} constraint.

\subsection{Governing equations of the unit cell}
\label{ssec:1D-chain-discrete-equilibrium-constitutive}

The work done by the microforces is
\begin{equation}
-2H\,\bm{e}_1\cdot\underbrace{\bigl(-\Delta\,\bm{e}_1+\bm{u}_{i-1}+\Delta\,\bm{t}_i\bigr)}_{\bm{C}'-\bm{C}}
+2H\,\bm{e}_1\cdot\underbrace{\bigl(\Delta\,\bm{e}_1+\bm{u}_i-\Delta\,\bm{t}_i\bigr)}_{\bm{G}'-\bm{G}}, 
\label{eq:energy-ext}
\end{equation}
where $\bm{C}'-\bm{C}$ and $\bm{G}'-\bm{G}$ are the displacement vectors of the points at which the microforces are applied.

The total potential energy of the unit cell is given by the elastic energy stored in the axial and rotational springs minus the work done by the end forces $\bm{N}_{i-1}^+$ and $\bm{N}_i^+$, end moments $M_{i-1}^+$ and $M_i^+$, and microforces $2H$. Using the kinematic relation introduced previously, it can be written as
\begin{multline}
W_i
=
\frac{k_r}{2}\,(\theta_{i+1}-\theta_i)^2
+
\frac{k_a}{2}\,d_i^{\,2}
+
M_{i-1}^+\,\theta_i
-
M_i^+\,\theta_{i+1} 
+
\bm{N}_{i-1}^+ \cdot \Bigl(\bm{u}_i+\ell_0\,\bm{e}_1-(\ell_0+d_i)\,\bm{t}_i\Bigr) \\
-
\bm{N}_i^+ \cdot \bm{u}_i 
+
2H\Bigl\{\ell_0 - 2\Delta - (\ell_0-2\Delta+d_i)\cos\theta_i\Bigr\}.
\label{eq:potential-dis}
\end{multline}
The stationarity of $W_i$ with respect to $\theta_i$, $\theta_{i+1}$, $\bm{u}_i$, and $d_i$ yields, respectively, the following equilibrium and constitutive equations
\begin{equation}
\begin{array}{lll}
\text{Moment balance} 
&
M_i^+ = M_{i-1}^+ + (\ell_0+d_i)\,\bm{N}_{i-1}^+\cdot\bm{n}_i + 2H(\ell_0-2\Delta+d_i)\sin\theta_i,
\\[3mm]
\text{Moment–rotation relation}
&
M_i^+ = k_r(\theta_{i+1}-\theta_i) ,
\\[3mm]
\text{Force balance}
&
\bm{N}_i^+ = \bm{N}_{i-1}^+ ,
\\[3mm]
\text{Axial constitutive relation}
&
\bm{N}_{i-1}^+\cdot\bm{t}_i=k_a d_i - 2H\cos\theta_i .
\end{array}
\label{eq:equilibrium-dis-simp}
\end{equation}
The microforces have zero resultant on each unit cell and therefore do not alter the transmitted resultant force between adjacent hinges.

\subsection{Analysis of the periodic discrete structure}
\label{ssec:1D-chain-discrete-analysis}
 
A chain consisting of $n$ elements and $n+1$ nodes is considered, with a total undeformed length $L = n \ell_0$. Up to a rigid-body translation, its deformed configuration is determined by the $n$ element rotations $\theta_i$ and the $n$ axial elongations $d_i$, for $i=1,\ldots,n$. Once these $2n$ kinematic variables are known, the $2(n+1)$ scalar components of the nodal displacements $\bm{u}_i$ (for $i=0,\ldots,n$) can be obtained recursively from the compatibility relation~\eqref{eq:ui-slaved}. The remaining rigid-body translation is specified by $\bm{u}_0 = \{u_0,v_0\}^{\mathsf T}$ and fixed by the boundary conditions.

The third relation in Eq.~\eqref{eq:equilibrium-dis-simp}$_3$ shows that the internal-force vector is independent of the node index $i$ and is therefore the same at every node of the chain. Denoting this constant vector by $\bm{N}$ leads to
\begin{equation}
\bm{N} = \bm{N}_0^+ = \bm{N}_i^+ = \bm{N}_n^+, \qquad i=1,\ldots,n-1, 
\end{equation}
with the corresponding horizontal and vertical components given by
\begin{equation}
N^x = \bm{N} \cdot \bm{e}_1,
\qquad
N^y = \bm{N} \cdot \bm{e}_2.
\end{equation}
The internal forces and moments at the endpoints are related to the corresponding external forces and moments through
\begin{equation}
\bm{N}_0^+=-\bm{F}_0, \qquad M_0^+=-C_0, \qquad \bm{N}_n^+=\bm{F}_n, \qquad M_n^+=C_n.
\end{equation}
The end forces $\bm{F}_0$ and $\bm{F}_n$, and the end moments $C_0$ and $C_n$ may be either prescribed external loads or unknown reactions, depending on the boundary conditions. Because the microforces have zero resultant, global force equilibrium requires
\begin{equation}
\bm{F}_0+\bm{F}_n=\bm{0}.
\end{equation}
For each element, the moment-balance equation and the axial constitutive relation, given by the first and fourth relations in Eq.~\eqref{eq:equilibrium-dis-simp}, respectively, are applied for $i=1,\ldots,n$. The $n-1$ internal moments at the hinges are expressed through the moment--rotation relation, given by the second relation in Eq.~\eqref{eq:equilibrium-dis-simp}, for $i=1,\ldots,n-1$. The two end moments $M_0^+$ and $M_n^+$ are prescribed by the endpoint loading conditions or treated as unknown reactions. 

When the end forces and moments are prescribed, eliminating the internal moments at the hinges leaves a system of $2n$ equations for the $n$ rotations and the $n$ axial elongations. If any endpoint forces or moments are unknown reactions, they must be included among the unknowns, and the corresponding kinematic boundary conditions must be added to close the system. Global moment equilibrium follows from summing the moment-balance equations for the $n$ elements and therefore need not be imposed separately.

\subsection{Bifurcation loads for a cantilever chain}
\label{ssec:1D-chain-discrete-cantilever}

\begin{figure}[H]
    \centering
    \includegraphics[width=1\linewidth]{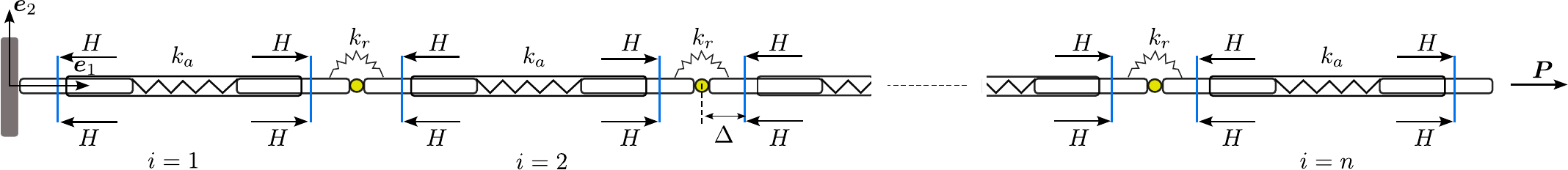}
    \caption{
    One-dimensional chain consisting of $n$ unit cells in a cantilever configuration. The chain is clamped at its left end and subjected to the periodic microforces, each of magnitude $2H$, and the horizontal force $P\,\bm{e}_1$ applied at its right end.
    }
    \label{fig:1D-chian-force-cantilever}
\end{figure}

Consider the cantilever configuration shown in Fig.~\ref{fig:1D-chian-force-cantilever}. The force applied at the right end is
\begin{equation}
\bm{F}_n = P\, \bm{e}_1,
\label{eq:discrete-cantilever-bcs-force}
\end{equation}
where $P > 0$ denotes tension and $P < 0$ denotes compression. The applied moment at the right end is zero, $C_n = 0$.

For this configuration, the governing equations are
\begin{equation}
\left\{
\begin{array}{@{}ll@{}}
k_a \ell_0\, \varepsilon_i - 2H \cos\theta_i - P \cos\theta_i = 0, & i = 1, \ldots, n, \\[3mm]
k_r (\theta_2 - \theta_1) + C_0 - P \ell_0 (1 + \varepsilon_1) \sin\theta_1 - 2H \ell_0 \left(1 - 2 \frac{\Delta}{\ell_0} + \varepsilon_1\right) \sin \theta_1 = 0, & \\[3mm]
k_r (\theta_{i+1} - 2 \theta_i + \theta_{i-1}) - P \ell_0 (1 + \varepsilon_i) \sin\theta_i - 2H \ell_0 \left(1 - 2 \frac{\Delta}{\ell_0} + \varepsilon_i\right) \sin\theta_i = 0, & i = 2, \ldots, n-1, \\[3mm]
k_r (-\theta_n + \theta_{n-1}) - P \ell_0 (1 + \varepsilon_n) \sin\theta_n - 2H \ell_0 \left(1 - 2 \frac{\Delta}{\ell_0} + \varepsilon_n\right) \sin\theta_n = 0. 
\end{array}
\right.
\label{eq:equilibrium-dis-detailed-cantilever-simp}
\end{equation}
The clamp at the left end imposes $\theta_1=0$. The reaction moment $C_0$ can be obtained by summing the $n$ moment--balance equations. Using $\theta_1 = 0$ gives
\begin{equation}
C_0 = \sum_{i=2}^n \ell_0\sin\theta_i \left( P(1+\varepsilon_i) + 2H\left(1-2\frac{\Delta}{\ell_0}+\varepsilon_i\right) \right),
\label{eq:C0}
\end{equation}
so that, once $C_0$ is evaluated from this equation, the moment-balance equation for the first element becomes redundant.

The system \eqref{eq:equilibrium-dis-detailed-cantilever-simp} admits the trivial solution
\begin{equation}
    \theta_i^{(0)} = 0, \qquad \varepsilon_i^{(0)} = \varepsilon^{(0)} = \frac{P+2H}{k_a\ell_0}, \qquad i=1,\ldots,n,
\end{equation}
which represents a uniformly stretched, but straight, configuration in equilibrium under the end force $P$ and the microforces $2H$. Mechanical admissibility requires the current length of the central axial spring to remain positive. Thus the admissibility condition is
\begin{equation}
\lambda^{(0)} = 1 + \varepsilon^{(0)} = 1 + \frac{P + 2H}{k_a\ell_0} > 2\frac{\Delta}{\ell_0}. \label{eq:condition-admissibility}
\end{equation}

Bifurcation occurs when the equations linearized about the straight configuration admit a nonzero rotational perturbation. At first order, the axial constitutive equations yield no perturbation in $\varepsilon_i$; hence, the strain remains equal to $\varepsilon^{(0)}$ up to first order.

The element rotations $\theta_i$ for $i=2,\ldots,n$ are collected into the $(n-1)$-dimensional vector
\begin{equation}
\bm{\theta} = \{\theta_2,\theta_3,\ldots,\theta_n\}^{\mathsf T}, 
\label{eq:theta-epsilon}
\end{equation}
so that the linearized moment-balance equations can be written as
\begin{equation}
k_r\, \bm{L}_{\text{DN}}\, \bm{\theta} + \ell_0 \mathcal{F}_d^{(0)}(P,H)\, \bm{\theta} = \bm{0},
\end{equation}
where
\begin{equation}
\mathcal{F}_d^{(0)}(P,H) = (P + 2H) \left(1 + \frac{P + 2H}{k_a\ell_0}\right) - 4H \frac{\Delta}{\ell_0} ,
\end{equation}
and $\bm{L}_{\text{DN}}$ is the $(n-1) \times (n-1)$ Dirichlet–Neumann discrete Laplacian matrix
\begin{equation}
\bm{L}_{\text{DN}}
=
\left[
\begin{array}{cccccc}
2  & -1 &        &        &        & \\
-1 & 2  & -1     &        &        & \\
   & -1 & 2      & -1     &        & \\
   &    & \ddots & \ddots & \ddots & \\
   &    &        & -1     & 2      & -1 \\
   &    &        &        & -1     & 1
\end{array}
\right] .
\label{eq:operator-L}
\end{equation}
Introducing 
\begin{equation}
\omega^2 = - \frac{\ell_0}{k_r} \mathcal{F}_d^{(0)}(P,H),
\end{equation}
the determination of the bifurcation loads reduces to the eigenvalue problem
\begin{equation}
\left( \bm{L}_{\text{DN}} - \omega^2 \bm{I}_{n-1} \right) \bm{\theta} = \bm{0}.
\label{eq:eq:equilibrium-dis-detailed-cantilever-comp-dimensionless}
\end{equation}
The eigenvalues of $\bm{L}_{\text{DN}}$ are
\begin{equation}
\omega_m^2 = 2 - 2 \cos \frac{(2 m - 1) \pi}{2 n - 1}  = 
4 \sin^2 \frac{(2 m - 1) \pi}{2 (2 n - 1)}, \quad 
m = 1, \ldots, n - 1,
\end{equation}
and the corresponding bifurcation condition is
\begin{equation}
(P + 2 H)\left(1 + \frac{P + 2 H}{k_a \ell_0}\right) - 4 H \frac{\Delta}{\ell_0} = 
- \frac{k_r}{\ell_0} \omega_m^2, \qquad m = 1, \ldots n-1.
\label{eq:cquadratic}
\end{equation}
Using $L = n \ell_0$ and assuming $H$ as a prescribed parameter, the two solutions of the quadratic equation in $P$ are
\begin{equation}
P_m^{\pm}(H) = -\frac{1}{2} 
\left\{
k_a\ell_0 + 4H \pm \sqrt{k_a\ell_0 \left(k_a\ell_0 - 4 n^2 \omega_m^2 \frac{k_r\ell_0}{L^2} + 16 H \frac{\Delta}{\ell_0}\right)}
\right\}, 
\label{eq:chaincr}
\end{equation}
for $m = 1, \ldots, n - 1$. For each mode $m$, the bifurcation condition defines a curve in the two-parameter loading space $(P,H)$. In the following, the signed microforce parameter $H$ is prescribed, and the bifurcation condition is solved for the critical values of the end force $P$, yielding the two branches $P_m^\pm(H)$. Real bifurcation loads exist only when the discriminant is nonnegative, and only roots satisfying $\lambda^{(0)} > 2\Delta/\ell_0$ are mechanically admissible.

The corresponding eigenmodes are 
\begin{equation}
\theta_i^{(m)} = \sin \left( \frac{(2m - 1) (i-1) \pi}{2n - 1} \right), \quad
i = 2, \ldots, n.
\end{equation}
Introducing dimensionless parameters
\begin{equation}
p = \frac{P}{k_a \ell_0}, \quad 
h = \frac{H}{k_a \ell_0}, \quad 
\delta = \frac{\Delta}{\ell_0}, \quad
\chi = \frac{k_r}{n^2 k_a \ell_0^2} = \frac{k_r}{k_a L^2}, 
\label{eq:dimensionless-arguments-discrete-a}
\end{equation}
the finite-chain bifurcation loads, for prescribed $h$, become
\footnote{
For a simply supported chain, the same calculation leads to the finite-chain bifurcation formula given above, with $h$ again treated as prescribed, but with the eigenvalues replaced by
$$
\omega_m^2 = 4 \sin^2 \frac{m\pi}{2n}, \quad m = 1, \ldots, n-1. 
$$
}
\begin{equation}
p_m^{\pm}(h) =
-\frac{1}{2} \left\{
1 + 4 h \pm \sqrt{1 - 4 \chi\, n^2 \omega_m^2 + 16 h \delta}
\right\}, \quad m = 1, 2, \ldots, n - 1.
\label{eq:p-critical-discrete}
\end{equation}
The continuum limit is obtained by letting $n = L/\ell_0$ tend to infinity while keeping $L$, $\delta$, $H$, $k_a\ell_0$, and $k_r\ell_0$ fixed. For each fixed mode number $m$, the limit yields
\begin{equation}
4 n^2 \omega_m^2 = 16 n^2 \sin^2 \frac{(2m - 1) \pi}{2(2n - 1)} \to (2m - 1)^2 \pi^2, \quad \text{as} \quad n \to \infty ,
\end{equation}
and, consequently, the two bifurcation branches converge to
\begin{equation}
P_m^{\pm}(H) = -\frac{1}{2} 
\left\{
k_a\ell_0 + 4H \pm \sqrt{k_a\ell_0 \left(k_a\ell_0 - (2m - 1)^2 \pi^2 \frac{k_r\ell_0}{L^2} + 16 H \frac{\Delta}{\ell_0}\right)}
\right\}, 
\label{eq:limitchain}
\end{equation}
which can equivalently be expressed in  the dimensionless form as
\begin{equation}
p_m^{\pm}(h) =
-\frac{1}{2} \left\{
1 + 4 h \pm \sqrt{1 - (2m - 1)^2 \pi^2 \chi + 16 h \delta}
\right\}.
\label{eq:p-critical-continuum}
\end{equation}
Equations~\eqref{eq:limitchain} and \eqref{eq:p-critical-discrete} show that the sequence of bifurcation loads is always finite, as real values exist only for a limited set of parameters and mode number $m$.

\section{Homogenization of the chain into equivalent elastic rod models}

Three homogenization strategies are introduced to derive nonlinear elastic rod models equivalent to the discrete chain presented in the previous section. The models differ in the treatment of the microscopic loading and the axial stiffness distribution, as summarized below.
\begin{itemize}
\item[i.] \textbf{Model (i): Axially and flexurally homogenized model incorporating the microforces as a state of prestress and an effective distribution of rotational springs (Sec.~\ref{subsec:Lagrangian}).} The discrete chain and its microscopic loading are homogenized simultaneously. A continuum total-potential-energy functional is obtained by accounting for both the elastic energy of the chain and the work done by the microforces. The equilibrium equations and boundary conditions are then derived using a variational approach. The same governing equations are obtained independently through direct homogenization of the discrete equilibrium equations. The resulting equivalent rod has uniform axial and flexural stiffnesses. The microforces no longer appear as explicitly applied external loads; instead, their effect is incorporated into two combined states of the internal forces generating a prestress state and a distribution of nonlinear rotational springs.
\item[ii.] \textbf{Model (ii): Axially and flexurally homogenized model subjected to a periodic distribution of discrete microforces (Sec.~\ref{ssec:impose-forces-hom-axially-flexurally-beam}).} The axial and flexural stiffnesses of the discrete chain are first homogenized without accounting for the microforces, yielding an equivalent rod with uniform elastic properties. The original periodic distribution of concentrated microforces is then applied to the rod. Thus, the microforces remain externally applied loads.
\item[iii.] \textbf{Model (iii): Flexurally homogenized model with piecewise-constant axial stiffness, subjected to a periodic distribution of discrete microforces (Sec.~\ref{ssec:impose-forces-hom-flexurally-beam}).} Only the flexural stiffness of the discrete chain is homogenized. The axial stiffness retains its original periodic, piecewise-constant distribution, alternating between axially deformable and inextensible segments. The resulting rod has uniform flexural stiffness and piecewise-constant axial stiffness. As in Model (ii), the original distribution of concentrated microforces is applied as an external loading.
\end{itemize}

\subsection{Model (i): Axially and flexurally homogenized model incorporating the microforces as prestress and distributed rotational springs}
\label{subsec:Lagrangian}

\subsubsection{Energy-based homogenization}

The total potential energy of the chain, consisting of $n$ discrete elements with total initial length $L=n\ell_0$, is obtained by summing the elastic energies stored in the axial and rotational springs and subtracting the work done by the external loads. Equivalently, it can be obtained by summing the potential energies of the individual unit cells given in Eq.~\eqref{eq:potential-dis}, 
\begin{multline}
\label{sheshe}
W = 
\sum_{i=1}^{n-1} \frac{k_r\ell_0}{2} \left(\frac{\theta_{i+1}-\theta_i}{\ell_0}\right)^2 \ell_0
+ \sum_{i=1}^{n} \frac{k_a\ell_0}{2} \left(\frac{d_i}{\ell_0}\right)^2 \ell_0 
- C_0 \theta_1 - C_n \theta_n - \bm{F}_0 \cdot \bm{u}_0 - \bm{F}_n \cdot \bm{u}_n \\[3mm]
- 2H \sum_{i=1}^{n} \bigg(2\frac{\Delta}{\ell_0}(1-\cos{\theta_i}) + (1+\varepsilon_i)\cos\theta_i - 1\bigg) \ell_0.
\end{multline}
The force applied at the right end of the chain is
\begin{equation}
\bm{F}_n = P \bm{e}_1 + V \bm{e}_2,
\label{eq:Fn}
\end{equation}
and global force equilibrium requires $\bm{F}_0 = - \bm{F}_n$. 

The discrete chain is homogenized by letting $\ell_0$ tend to zero while keeping the total length $L$ and the ratio $\Delta/\ell_0$ fixed, so that $n = L/\ell_0$ tends to infinity. 
\begin{equation}
\begin{array}{lll}
\displaystyle \theta'(s) = \lim_{\ell_0 \to 0} \frac{\theta_{i+1} - \theta_i}{\ell_0}, & 
\displaystyle \varepsilon(s) = \lim_{\ell_0 \to 0} \frac{d_i}{\ell_0} = \lambda(s) - 1, & 
\text{curvature and axial strain} \\[5mm]
\displaystyle \bm{t}(s) = \lim_{\ell_0 \to 0} \bm{t}_i, & 
\displaystyle \bm{n}(s) = \lim_{\ell_0 \to 0} \bm{n}_i, & 
\text{unit tangent and normal} 
\end{array}
\label{eq:limit-kinematic-material}
\end{equation}
where $\lambda$ is the stretch. Moreover, a nondegenerate continuum limit with finite axial and flexural stiffnesses $EA$ and $EI$ requires the discrete spring stiffnesses $k_a$ and $k_r$ to scale inversely with $\ell_0$, such that
\begin{equation}
\begin{array}{lll}
\displaystyle EI = k_r \ell_0, & 
\displaystyle EA = k_a \ell_0. & 
\text{flexural and axial stiffness}
\end{array}
\label{eq:limit-kinematic-material2}
\end{equation}
The bending moment and elastic axial force are given by
\begin{equation}
\begin{array}{lll}
\displaystyle M(s) = EI\, \theta'(s), & 
\displaystyle N(s) = EA\, \varepsilon(s). & 
\text{}
\end{array}
\end{equation}
In the continuum limit, the discrete sums become integrals, and the total potential energy functional assumes the form
\begin{multline}
W = \int_0^L \frac{EI}{2}(\theta')^2\,ds + \int_0^L \frac{EA}{2}\varepsilon^2\, ds 
- C_0\,\theta(0) - C_n\,\theta(L) - \bm{F}_n \cdot \bigl(\bm{u}(L)-\bm{u}(0)\bigr) \\[3mm]
- 2H \int_0^L \bigg(2\frac{\Delta}{\ell_0}(1-\cos\theta) + (1+\varepsilon)\cos\theta -1\bigg)ds.
\end{multline}
The continuum limit of the discrete kinematic constraint~\eqref{eq:ui-slaved} is
\begin{equation}
\bm{u}' =  \lim_{\ell_0 \to 0} \frac{\bm{u}_i-\bm{u}_{i-1}}{\ell_0} = 
(1+\varepsilon)\bm{t}-\bm{e}_1,
\end{equation}
while writing the displacement vector as $\bm{u}(s) = u(s)\, \bm{e}_1 + v(s)\, \bm{e}_2$ leads to 
\begin{equation}
u' = \lambda\cos\theta - 1,
\quad
v' = \lambda\sin\theta.
\label{eq:uprime-vprime}
\end{equation}
The equivalent rod is unshearable, imposing the constraint
\begin{equation}
\gamma = v'\cos\theta - (1 + u')\sin\theta = 0, 
\label{eq:unshearable}
\end{equation}
and the stretch and axial strain are
\begin{equation}
\lambda = (1 + u')\cos\theta + v'\sin\theta, \qquad \varepsilon = \lambda - 1.
\end{equation}
Using the condition $(1 + \varepsilon) \cos\theta - 1 = u'$, the augmented total potential energy functional is expressed as
\begin{multline}
\hat{W} = \int_0^L \frac{EI}{2} (\theta')^2\, ds + \int_0^L \frac{EA}{2} \varepsilon^2\, ds
- C_0 \,\theta(0) - C_n \,\theta(L) 
- \bm{F}_n \cdot (\bm{u}(L) - \bm{u}(0)) \\[3mm] 
- \int_0^L 2 H \left\{ 2\frac{\Delta}{\ell_0}(1 - \cos \theta) + u' \right\}\, ds 
+ \int_0^L T \gamma\, ds ,
\label{eq:potential-hom-agumented}
\end{multline}
where $T$ is a Lagrange multiplier enforcing the unshearability constraint $\gamma = 0$, which is subsequently identified with the internal shear-force resultant. 

The variations of the axial strain and shear measure are
\begin{equation}
\begin{aligned}
\delta{\varepsilon} &= 
\Big[v'\cos\theta - (1 + u')\sin\theta\Big] \delta{\theta} + \cos\theta\, \delta{u'} + \sin\theta\, \delta{v'} , \\[3mm]
\delta{\gamma} &= 
- \Big[v'\, \sin\theta + (1 + u')\cos\theta\Big] \delta{\theta} 
- \sin\theta\, \delta{u'} + \cos\theta\, \delta{v'}.
\end{aligned}
\end{equation}
The first variation of the augmented total potential energy can be written as  
\begin{multline}
\delta{\hat{W}} = 
\int_0^L EI\, \theta'\, \delta{\theta'}\, ds + \int_0^L EA\, \varepsilon\, \delta{\varepsilon}\, ds - \int_0^L 2H \left\{ 2\frac{\Delta}{\ell_0}\sin\theta\, \delta{\theta} + \delta{u'} \right\}\, ds 
+ \int_0^L T\, \delta{\gamma}\, ds \\[3mm]
+ \int_0^L \gamma\, \delta{T}\, ds - C_0\, \delta{\theta(0)} - C_n\, \delta{\theta(L)} + P\, \delta{u(0)} + V\, \delta{v(0)} - P\, \delta{ u(L)} - V\, \delta{v(L)}.
\label{eq:potential-hom-variation}
\end{multline}
After integration by parts and collection of the independent variations, one obtains
\begin{multline}
\delta{\hat{W}}
= 
\int_0^L \Big( - EI\, \theta'' + EA\, \varepsilon\, \gamma - T\, \lambda - 4 H \frac{\Delta}{\ell_0} \sin\theta \Big) \delta{\theta}\, ds + \int_0^L \gamma\, \delta{T}\, ds \\[3mm]
+ \int_0^L \Big( - EA (\varepsilon \cos\theta)' + (T \sin\theta)' \Big) \delta{u}\, ds 
+ \int_0^L \Big( - EA (\varepsilon \sin\theta)' - (T \cos\theta)' \Big) \delta{v}\, ds \\[3mm]
+ \Big[ \bigl( EA\, \varepsilon\, \cos\theta - P - T \sin\theta - 2H \bigr)\delta{u} \Big]_{s=L}
- \Big[ \bigl( EA\, \varepsilon\, \cos\theta - P - T \sin\theta - 2H \bigr)\delta{u} \Big]_{s=0} \\[3mm]
+ \Big[ \bigl( EA\, \varepsilon\, \sin\theta - V + T \cos\theta \bigr) \delta{v} \Big]_{s=L} 
- \Big[ \bigl( EA\, \varepsilon\, \sin \theta - V + T \cos\theta \bigr) \delta{v} \Big]_{s=0} \\[3mm]
+ \Big[ \bigl( EI\, \theta' - C_n\bigr)\delta \theta \Big]_{s=L}
- \Big[ \bigl( EI\, \theta' + C_0\bigr)\delta \theta \Big]_{s=0} .
\label{eq:potential-hom-variation-interation-simp}
\end{multline}
Stationarity with respect to $u$ and $v$ gives the force-equilibrium equations
\begin{equation}
(EA\, \varepsilon \cos\theta - T \sin\theta)' = 0,
\qquad
(EA\, \varepsilon \sin\theta + T \cos\theta)' = 0.
\end{equation}
The associated natural boundary conditions are
\begin{equation}
\Big[EA\, \varepsilon \cos\theta - T \sin\theta\Big]_{s=0,L} = P + 2H,
\qquad
\Big[EA\, \varepsilon \sin\theta + T \cos\theta\Big]_{s=0,L} = V, 
\end{equation}
which can be solved for $EA\, \varepsilon$ and $T$ to yield
\begin{equation}
EA\, \varepsilon = (P + 2H)\cos\theta + V\sin\theta,
\qquad
T = V\cos\theta - (P + 2H)\sin\theta.
\label{eq:axial-shear-resultants-model-i}
\end{equation}
Thus, $N = EA\,\varepsilon$ is the axial-force resultant in the direction of the rod tangent, while $T$ is the transverse shear-force resultant in the direction of the rod normal. The latter acts as the reaction force enforcing the unshearability constraint. Stationarity with respect to the Lagrange multiplier $T$ recovers the unshearability constraint $\gamma = 0$.

Stationarity with respect to $\theta$ gives the rotational equilibrium equation 
\begin{equation}
\label{tagliazzo}
EI \theta''(s) + T(s) \lambda(s) + 4H \frac{\Delta}{\ell_0} \sin\theta(s) = 0.
\end{equation}
The last term in Eq.~(\ref{tagliazzo}) represents a rotation-dependent distributed moment generated by the homogenization of the microforces. In the equivalent continuum rod model, this contribution can be interpreted as the action of nonlinear rotational springs continuously distributed along the rod axis, providing the rotational analog of a Winkler foundation. Remarkably, for positive microforces, $H>0$, and $\theta \in (-\pi/2,\pi/2)$, 
{\it 
the distributed springs exhibit negative tangent stiffness -- a response that would be difficult, if not impossible, to realize by conventional mechanical means.
}
More precisely, the moment exerted by the distributed springs on the rod is
\begin{equation}
c_H(\theta) = 4 H \frac{\Delta}{\ell_0} \sin\theta,
\end{equation}
which varies sinusoidally with the rotation and vanishes in the straight configuration, $\theta = 0$. Adopting the convention that a spring with positive stiffness exerts a restoring moment opposite to the rotation, the tangent stiffness of the distributed springs is
\begin{equation}
k_H^{\text{tan}}(\theta) = -\frac{d c_H}{d\theta} = -4 H \frac{\Delta}{\ell_0} \cos\theta.
\end{equation}

Using the condition $\lambda = 1 + \varepsilon$ and the expression for $T$, the rotational equilibrium equation becomes
\begin{equation}
\theta''(s) - \frac{1}{EI} \left\{
(P + 2 H)\bigl(1 + \varepsilon(s)\bigr) - 4 H \frac{\Delta}{\ell_0}
\right\} \sin\theta(s)
+ \frac{V}{EI} \bigl(1 + \varepsilon(s)\bigr) \cos\theta(s) = 0,
\label{eq:equi-hom-sim}
\end{equation}
where the axial strain is given by
\begin{equation}
\varepsilon(s) = \frac{(P + 2 H)\cos\theta(s) + V\sin\theta(s)}{EA}. 
\label{eq:axialeps}
\end{equation}
The natural boundary conditions for the bending moment follow from the rotational boundary terms:
\begin{equation}
M(0) = EI\, \theta'(0) = -C_0, \qquad M(L) = EI\, \theta'(L) = C_n.
\end{equation}
At each endpoint, either a kinematic variable or its conjugate force/moment resultant may be prescribed. The corresponding pairs of boundary conditions are:
\begin{itemize}
\item Horizontal displacement $u$ or horizontal force $N^x =N\, \cos\theta - T\, \sin\theta = P + 2H$.
\item Vertical displacement $v$ or vertical force $N^y = N\, \sin\theta + T\, \cos\theta = V$.
\item Rotation $\theta$ or bending moment $M = EI\, \theta'$.
\end{itemize}
The rotational equilibrium equation, together with the appropriate boundary conditions, defines a nonlinear boundary-value problem for $\theta(s)$. Once $\theta(s)$ is determined, the axial strain and transverse shear-force resultant follow from Eqs.~\eqref{eq:axialeps} and \eqref{eq:axial-shear-resultants-model-i}$_2$, respectively. The kinematic equations~\eqref{eq:uprime-vprime} can then be integrated to determine the displacement fields $u(s)$ and $v(s)$.

The main result of the homogenization is that the microscale forces produce two different effects in the equivalent rod:
\begin{enumerate}
\item \textbf{An effective axial prestress}: The microforce term $2H$ adds to the applied end force $P$, giving an effective axial prestress
\begin{equation}
\label{prrre}
N^x = P+2H.
\end{equation}
Thus, although the horizontal external force applied at the right end of the original discrete chain is $P$, the equivalent rod is subjected to the effective horizontal end force $P+2H$. 
\item \textbf{An effective constraint corresponding to  rotational springs distributed along the rod axis}: In addition to contributing to the prestress, the microforces also generate the nonlinear term 
\begin{equation}
\label{prrro}
4H(\Delta/\ell_0)\sin\theta,
\end{equation}
which enters directly into the rotational equilibrium equation (\ref{tagliazzo}) and can be interpreted as a distribution of nonlinear elastic springs. These springs have negative tangent stiffness when $H>0$, so that they play a destabilizing role in Eq.~(\ref{eq:equi-hom-sim}).
\end{enumerate}
The two contributions (\ref{prrre}) and (\ref{prrro}) combine in the rotational equilibrium equation in Eq.~(\ref{eq:equi-hom-sim}) through the effective load coefficient
\begin{equation}
\mathcal{F}_{I}(s) = (P + 2 H)\bigl(1 + \varepsilon(s)\bigr) - 4 H \frac{\Delta}{\ell_0},
\end{equation}
which should not be interpreted as an axial prestress, because the latter remains $N^x = P + 2H$, as established by the natural boundary condition. Instead, $\mathcal{F}_I$ is an effective geometric-stiffness parameter that combines the moment generated by the axial prestress along the stretched rod with an additional moment generated by the microforces and equivalent to a moment produced by nonlinear rotational springs distributed along the rod axis. Because of their different signs, the two contributions of $H$ have opposite effects on the stability of the rod.

For $V=0$, the straight configuration is characterized by $\varepsilon^{(0)}=(P+2H)/EA$, and the effective load coefficient reduces to
\begin{equation}
\mathcal{F}_I^{(0)}=(P+2H)\left(1+\frac{P+2H}{EA}\right)-4H\frac{\Delta}{\ell_0},
\label{eq:G0}
\end{equation}
which determines the geometric stiffness of the linearized rotational problem and the onset of bifurcation.

Homogenization, therefore, transforms the periodic microscale loading into an effective uniform prestress and a rotational spring distribution. This result fundamentally distinguishes the proposed scheme from models in which the microforces remain explicitly applied external loads.

\subsubsection{Direct homogenization from equilibrium}

An independent derivation of the rotational equilibrium equation can be obtained by performing the continuum limit of the discrete equilibrium equations. For simplicity, the superscript `+' is omitted in this subsection.

Substituting the moment/rotation relation, Eq.~\eqref{eq:equilibrium-dis-simp}$_2$, into the moment-balance equation~\eqref{eq:equilibrium-dis-simp}$_1$, and writing the internal-force vector as $\bm{N}_{i-1} = N^x_{i-1} \bm{e}_1 + N^y_{i-1} \bm{e}_2$, the following equilibrium equation is obtained:
\begin{equation}
-(M_i - M_{i-1}) + \left\{(N^x_{i-1} + 2H)(\ell_0 + d_i) - 4H\Delta\right\}\sin\theta_i - N^y_{i-1}(\ell_0 + d_i)\cos\theta_i = 0.
\label{eq:equi-dis-sim}
\end{equation}
The force-balance equation~\eqref{eq:equilibrium-dis-simp}$_3$ shows that the horizontal and vertical components of the internal-force vector transmitted between adjacent cells are independent of $i$. The boundary conditions therefore give $N_i^x = P$, and $N_i^y = V$ for $i = 0, \ldots, n$. Here, $P$ and $V$ are the horizontal and vertical components of the force applied at the right end of the discrete chain.

The axial constitutive equation~\eqref{eq:equilibrium-dis-simp}$_4$ can then be written as 
\begin{equation}
k_a d_i = (P + 2H)\cos\theta_i + V\sin\theta_i, 
\end{equation}
a relation that is substituted into the discrete moment-balance equation \eqref{eq:equi-dis-sim}. The resulting equation is divided by $\ell_0$, and the continuum limit is performed using the identifications (\ref{eq:limit-kinematic-material2}). The result is 
\begin{equation}
M'(s) - \left\{(P + 2H)\bigl(1 + \varepsilon(s)\bigr) - 4H\frac{\Delta}{\ell_0}\right\}\sin\theta(s) + V\bigl(1 + \varepsilon(s)\bigr)\cos\theta(s) = 0,
\label{eq:equi-hom}
\end{equation}
where $\varepsilon(s)$ is given by Eq.~\eqref{eq:axialeps}. Using the constitutive relation $M(s) = EI\, \theta'(s)$, Eq.~\eqref{eq:equi-hom-sim} is recovered.

\subsubsection{Bifurcation loads for the equivalent rod in a  cantilever configuration}

The cantilever discrete chain analyzed in Sec.~\ref{ssec:1D-chain-discrete-cantilever} is now considered. The corresponding homogenized equivalent rod, resulting from the homogenization approach developed in Sec.~\ref{subsec:Lagrangian}, is also a cantilever, but subjected to the effective horizontal end force $(P + 2H)\, \bm{e}_1$, and to the additional microforce-induced contribution appearing in the rotational equilibrium equation.

The straight configuration is characterized by
\begin{equation}
\theta(s) = 0, \qquad \varepsilon(s) = \varepsilon^{(0)}(s) = \frac{P + 2H}{EA}.
\end{equation}
A linearization of the equilibrium and axial constitutive equations about the uniformly prestressed state yields 
\begin{equation}
\theta''(s) - \frac{\mathcal{F}_I^{(0)}}{EI} \theta(s) = 0,
\label{eq:rotation}
\end{equation}
where $\mathcal{F}_I^{(0)}$ is the effective geometric-stiffness parameter introduced in Eq.~\eqref{eq:G0}. Defining $\alpha^2 = -\mathcal{F}_I^{(0)}/EI$, the bifurcation problem reduces to 
\begin{equation}
\theta''(s) + \alpha^2 \theta(s) = 0.
\label{eq:linearized}
\end{equation}
For a rod clamped at $s = 0$ and subject to a dead horizontal force at $s = L$, which does not apply any bending moment, the boundary conditions are
\begin{equation}
\theta(0) = 0, \quad M(L) = EI\, \theta'(L) = 0.
\label{eq:bc}
\end{equation}
The corresponding eigenmodes are
\begin{equation}
\theta_m(s) = A_m \sin \frac{(2m - 1) \pi s}{2L}, \quad m = 1, 2, \dots,
\end{equation}
and the eigenvalues satisfy
\begin{equation}
\alpha_m^2 L^2 = \frac{(2m - 1)^2 \pi^2}{4}, \quad m = 1, 2, \dots
\end{equation}
Consequently, the bifurcation condition for the $m$-th mode is
\begin{equation}
\mathcal{F}_I^{(0)}(P,H) = - \frac{(2m - 1)^2 \pi^2}{4} \frac{EI}{L^2}, \quad m = 1, 2, \dots
\label{eq:quadratic}
\end{equation}
For each mode $m$, Eq.~(\ref{eq:quadratic}) defines a critical curve in the two-parameter loading space $(P,H)$. If $H$ is prescribed, solving the equation for $P$ recovers the two branches of critical loads previously obtained by taking the continuum limit of the discrete chain, that is Eq.~\eqref{eq:limitchain} with the identifications $k_a \ell_0 = EA$ and $k_r \ell_0 = EI$. The explicit expressions and their dimensionless forms are therefore not repeated here. This agreement confirms that the homogenized Model (i) correctly reproduces the continuum limit of the discrete system.

\subsubsection{Instability of an equivalent rod with null axial resultant}
\label{fifa}

Consider the cantilever chain analyzed in Sec.~\ref{ssec:1D-chain-discrete-cantilever}, without vertical force, under the special loading condition
\begin{equation}
P + 2H = 0.
\label{eq:zero-effective-axial-force}
\end{equation}
The macroscopic end force and the axial contribution of the microforces therefore cancel each other. Indeed, the axial constitutive equation of the discrete chain reduces to
\begin{equation}
k_a d_i = (P + 2H)\cos\theta_i = 0,
\end{equation}
so that the axial springs remain unloaded and the chain preserves its original length.

The corresponding equivalent rod has zero horizontal resultant,
\begin{equation}
N^x(s) = P + 2H = 0.
\end{equation}
Moreover, the exact axial constitutive relation gives
\begin{equation}
EA\,\varepsilon(s) = (P + 2H)\cos\theta(s) = 0,
\end{equation}
and hence, the axial strain vanishes in both the straight and bifurcated configurations:
\begin{equation}
\varepsilon(s) = 0.
\end{equation}
Nevertheless, the system is not load-free: the macroscopic force $P = -2H$ and the microforces are individually nonzero. Although their contributions to the axial resultant cancel, the effect of the microforces remains present as an effective distribution of rotational springs. Under condition~\eqref{eq:zero-effective-axial-force}, the nonlinear rotational equilibrium equation~\eqref{eq:equi-hom-sim} reduces to
\begin{equation}
\theta''(s) + \frac{4H}{EI}\frac{\Delta}{\ell_0}\sin\theta(s) = 0.
\label{eq:zero-axial-resultant-nonlinear}
\end{equation}
Equation~(\ref{eq:zero-axial-resultant-nonlinear}) can also be directly determined from the reduced potential energy,
\begin{equation}
W = \int_0^L 
\left[
\frac{EI}{2}\bigl(\theta'(s)\bigr)^2
- 4H\frac{\Delta}{\ell_0}\bigl(1-\cos\theta(s)\bigr)
\right]\, ds.
\label{nonnaabelarda}
\end{equation}
The Taylor expansion of Eq.~(\ref{nonnaabelarda}) about the straight configuration, up to and including quadratic terms, is
\begin{equation}
W^{(2)} = \frac{1}{2} \int_0^L
\left[
EI\bigl(\theta'(s)\bigr)^2 - 4H\frac{\Delta}{\ell_0}\theta^2(s)
\right]\, ds.
\end{equation}
Thus, for $H > 0$, the microforce contribution is destabilizing and can overcome the bending stiffness, even though the axial resultant is zero.

Linearization of Eq.~\eqref{eq:zero-axial-resultant-nonlinear} about the straight configuration gives
\begin{equation}
\theta''(s) + \frac{4H}{EI}\frac{\Delta}{\ell_0}\theta(s) = 0,
\end{equation}
together with the cantilever boundary conditions
\begin{equation}
\theta(0)=0, \qquad \theta'(L)=0.
\end{equation}
A nontrivial solution exists when
\begin{equation}
\frac{4H}{EI}\frac{\Delta}{\ell_0} = \frac{(2m-1)^2\pi^2}{4L^2},
\qquad
m=1,2,\ldots 
\end{equation}
The corresponding critical values of the microforce parameter are therefore
\begin{equation}
H_m^{\mathrm{cr}} = \frac{(2m-1)^2\pi^2}{16L^2} \frac{EI}{\Delta/\ell_0},
\qquad
m=1,2,\ldots 
\label{eq:Hcr}
\end{equation}
The corresponding value of the end force applied to the original discrete chain is 
\begin{equation} 
P_m^ {\mathrm{cr}} = -2H_m^{\mathrm{cr}}. 
\end{equation} 
By contrast, the effective horizontal end force acting on the equivalent rod is 
\begin{equation} 
\bigl(P + 2H\bigr)_{\mathrm{cr}} = 0, 
\end{equation} 
therefore, the straight equivalent rod can undergo an Euler-type bifurcation despite being subjected to zero axial end force and having zero axial strain. Unlike conventional Euler buckling, this instability is not produced by a compressive axial resultant, but by an equivalent distribution of rotational springs with negative stiffness. The effect disappears when $\Delta = 0$, confirming that it originates from the arms of the microforces, generating a distributed moment.

\subsection{Model (ii): Periodic microforces applied to an axially and flexurally homogenized rod}
\label{ssec:impose-forces-hom-axially-flexurally-beam}

\subsubsection{Governing equation with piecewise-constant coefficient}

A model is developed here in which the elastic properties of the discrete chain introduced in Sec.~\ref{sec:1D-chain-discrete} are first homogenized without accounting for the microforces. The resulting equivalent rod has uniform axial and flexural stiffnesses, $EA$ and $EI$, respectively. The original periodic distribution of microforces is then applied to this rod as an external loading.

In the absence of the microforces, an axially extensible rod, subjected to the constant end force $P\,\bm{e}_1+V\,\bm{e}_2$, satisfies
\begin{equation}
\theta''(s) - \frac{P}{EI} (1 + \varepsilon(s))\sin\theta(s) + \frac{V}{EI}  (1 + \varepsilon(s))\cos\theta(s) = 0. 
\label{eq:equilibrium-without-f}
\end{equation}
%

\begin{figure}[h]
    \centering
    \includegraphics[width=0.8\linewidth]{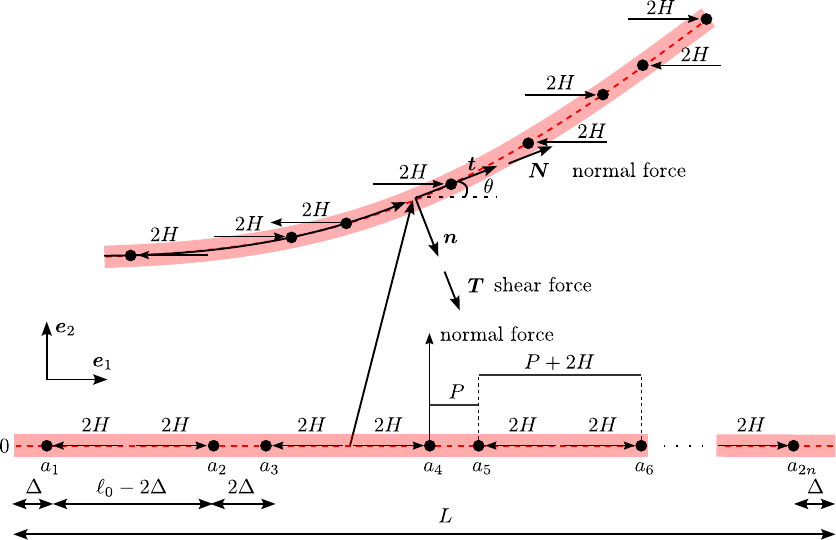}
    \caption{
    Lower part: periodic distribution of concentrated microforces along a
    one-dimensional rod of length $L$. The signed forces $\pm 2H\,\bm{e}_1$ are applied at the positions $a_i$, with $i=1,\ldots,2n$. Upper part: deformation of the rod under the microforces and the forces applied at its ends, which are not shown. Because the microforces are dead loads, they remain horizontal in the deformed configuration. In region~\#1, the horizontal internal resultant is $P+2H$, whereas in region~\#2 it is $P$.
    }
    \label{fig:post-hom-micro-force}
\end{figure}

On this rod, as illustrated in Fig.~\ref{fig:post-hom-micro-force}, the microforces are applied at the positions
\begin{equation}
a_{2k - 1} = (k - 1)\ell_0 + \Delta, \qquad 
a_{2k} = k\,\ell_0 - \Delta,
\qquad k = 1, 2, \ldots, n, 
\label{eq:position-a}
\end{equation}
and their distribution can be represented using the Dirac delta function, denoted here as $\delta_D$
\begin{equation}
\bm{q}(s) = 
2H\, \bm{e}_1 \sum_{k = 1}^{n} \Big\{ - \delta_D \left(s - \bigl((k - 1)\ell_0 + \Delta\bigr)\right) + \delta_D \left(s - \bigl(k\,\ell_0 - \Delta\bigr)\right) \Big\}.
\label{eq:q-explict}
\end{equation}
With the orientation of $\bm{n}$ adopted in Eq.~\eqref{eq:unit-tangent-normal}, the internal force resultant is written as
\begin{equation}
\bm{R}(s) = N(s)\,\bm{t}(s) - T(s)\,\bm{n}(s),
\end{equation}
where $N$ and $T$ are the axial and shear resultants, respectively. The force- and moment-equilibrium equations are
\begin{equation}
\bm{R}'(s) = -\bm{q}(s), 
\quad
M'(s) = -\lambda(s)\, T(s).
\label{eq:A-M}
\end{equation}
Because the distributed microforces have no vertical component, the Cartesian components of the internal resultant satisfy
\begin{equation}
(N^x)' = (N\cos\theta - T\sin\theta)' = - \bm{q} \cdot \bm{e}_1,
\quad 
(N^y)' = (N\sin\theta + T\cos\theta)' = 0.
\label{eq:equilibrium-components}
\end{equation}
Integrating Eq.~\eqref{eq:equilibrium-components} and imposing the end force $P\,\bm{e}_1 + V\,\bm{e}_2$ at $s = L$ gives
\begin{equation}
N^x = N \cos\theta - T \sin\theta  =  P + 2H\, \mathcal{I}(s),
\quad
N^y = N \sin\theta + T \cos\theta =  V, 
\label{eq:N-T-related}
\end{equation}
where
\begin{equation}
\mathcal{I}(s) = \sum_{k = 1}^{n} \Big\{ \Theta\bigl(s - ((k - 1)\ell_0 + \Delta)\bigr) - \Theta\bigl(s - (k\,\ell_0 - \Delta)\bigr) \Big\}, 
\label{eq:alpha}
\end{equation}
where $\Theta$ is the Heaviside step function. Thus, $\mathcal{I}=1$ in the central interval of each cell and $\mathcal{I}=0$ in its two outer intervals. Solving Eq.~\eqref{eq:N-T-related} for $N$ and $T$ yields
\begin{equation}
\begin{aligned}
N(s) &= \bigl[P+2H\mathcal{I}(s)\bigr]\cos\theta(s) + V\sin\theta(s), \\[2mm]
T(s) &= -\bigl[P+2H\mathcal{I}(s)\bigr]\sin\theta(s) + V\cos\theta(s).
\end{aligned}
\label{eq:N-T-post-hom}
\end{equation}
Using $M = EI\, \theta'$, $\lambda = 1 + \varepsilon$, and the second equation in Eq.~\eqref{eq:A-M}, the rotational equilibrium equation becomes
\begin{equation}
\theta''(s) - \frac{P + 2H\, \mathcal{I}(s)}{EI} (1+\varepsilon(s)) \sin\theta(s) + \frac{V}{EI} (1+\varepsilon(s))\cos\theta(s) = 0.
\label{eq:equil-post-hom-micro-force}
\end{equation}
The axial strain is piecewise-defined by
\begin{equation}
\varepsilon(s) = 
\frac{\bigl[P + 2H\, \mathcal{I}(s)\bigr]\cos\theta(s) + V \sin\theta(s)}{EA} . 
\label{eq:epsilon-post-hom-micro-force}
\end{equation}

\subsubsection{Bifurcation loads for a cantilever rod with a periodic distribution of microforces} 
\label{zorro}

A cantilever rod is analyzed, subjected to a horizontal dead force $P\,\bm{e}_1$ at its right end and to the periodic distribution of microforces shown in Fig.~\ref{fig:post-hom-micro-force}. Setting $V=0$, the straight equilibrium configuration is piecewise prestrained, with
\begin{equation}
\varepsilon^{(0)}(s) = \frac{P+2H\, \mathcal{I}(s)}{EA}.
\end{equation}
Linearization of Eqs.~\eqref{eq:equil-post-hom-micro-force} and \eqref{eq:epsilon-post-hom-micro-force} about this configuration gives
\begin{equation}
\theta''(s) + \beta^2(s) \theta(s) = 0, 
\label{eq:equilibrium-post-hom-micro-force}
\end{equation}
where 
\begin{equation}
\beta^2(s) = -\frac{P+2H\, \mathcal{I}(s)}{EI}\left(1+\frac{P+2H\, \mathcal{I}(s)}{EA}\right).
\end{equation}
Within each unit cell, the coefficient $\beta$ is a piecewise constant function of $s$:
\begin{equation}
\beta^2(s) =
\begin{cases}
\beta_1^2 = - \dfrac{P + 2H}{EI} \left(1 + \dfrac{P + 2H}{EA}\right),
& (k-1)\ell_0 + \Delta < s < k\ell_0 - \Delta; ~ k=1,2,\ldots,n,\\[5mm]
\beta_2^2 = - \dfrac{P}{EI} \left(1 + \dfrac{P}{EA}\right),
& \text{otherwise}.
\end{cases}
\label{eq:betaII}
\end{equation}
Thus, each cell contains a central interval of length $\ell_0-2\Delta$, governed by $\beta_1$, and two outer intervals of length $\Delta$, governed by $\beta_2$. If either $\beta_1^2$ or $\beta_2^2$ is negative, the following transfer matrices are understood through their equivalent hyperbolic form. 

To address the Euler-type bifurcation problem, the transfer matrix method (TMM) is employed, following an approach similar to that used in~\cite{fu2024wrinkling}. The state vector, collecting the rotation $\theta(s)$ and its derivative $\theta'(s)$, is defined as
\begin{equation}
\bm{y}(s)=
\begin{Bmatrix}
\theta(s) \\[2mm]
\theta'(s)
\end{Bmatrix}.
\label{eq:y-defined}
\end{equation}
For a segment of length $\ell$ governed by a constant value of $\beta$, the transfer matrix is
\begin{equation}
\bm{\Gamma}(\beta,\ell)=
\begin{pmatrix}
\cos(\beta\,\ell) & \dfrac{\sin(\beta\,\ell)}{\beta} \\[3mm]
-\beta \sin(\beta\,\ell) & \cos(\beta\,\ell)
\end{pmatrix}, 
\label{eq:Transfer-Matrix}
\end{equation}
whose derivation is deferred to  Appendix~\ref{sub_app:Transfer-Matrix-unit-cell}. If $\beta$ vanishes, the transfer matrix is understood in the limiting form obtained by continuity, namely
\begin{equation}
\bm{\Gamma}(\beta,\ell)=
\begin{pmatrix}
1 & \ell \\[3mm]
0 & 1
\end{pmatrix}.
\end{equation}
The transfer matrix of one unit cell is therefore
\begin{equation}
\bm{y}(\ell_0) = 
\widetilde{\bm{\Gamma}} \,
\bm{y}(0), 
\label{eq:unit-cell-transfer}
\end{equation}
where 
\begin{equation}
\widetilde{\bm{\Gamma}} = 
\bm{\Gamma}(\beta_2,\Delta)\,
\bm{\Gamma}(\beta_1,\ell_0-2\Delta)\,
\bm{\Gamma}(\beta_2,\Delta), 
\label{eq:tildegamma}
\end{equation}
with components 
\begin{equation}
\left\{
\begin{array}{l}
\widetilde{\Gamma}_{11} = \widetilde{\Gamma}_{22} = 
c_1(c_2^2-s_2^2) - c_2s_1s_2 
\left(\dfrac{\beta_1}{\beta_2} + \dfrac{\beta_2}{\beta_1}\right), \\[5mm]
\widetilde{\Gamma}_{12} = 
\dfrac{2c_1c_2s_2}{\beta_2} + s_1
\left(\dfrac{c_2^2}{\beta_1} - \dfrac{\beta_1}{\beta_2^2}s_2^2\right), \\[5mm]
\widetilde{\Gamma}_{21} = -2\beta_2c_1c_2s_2 + s_1
\left(\dfrac{\beta_2^2}{\beta_1}s_2^2 - \beta_1 c_2^2\right),
\end{array}
\right.
\end{equation}
in which
\begin{equation}
c_1 = \cos\Big(\beta_1(\ell_0 - 2\Delta)\Big),
\quad 
c_2 = \cos(\beta_2 \Delta),
\quad 
s_1 = \sin\Big(\beta_1(\ell_0 - 2\Delta)\Big), 
\quad 
s_2 = \sin(\beta_2 \Delta).
\end{equation}
For a rod consisting of $n$ identical cells, with $L = n\ell_0$, the state
vectors at its ends are related by
\begin{equation}
\bm{y}(L) = \widetilde{\bm{\Gamma}}^{n}\, \bm{y}(0), 
\label{eq:y-transfered}
\end{equation}
where, as shown in Appendix~\ref{sub_app:Transfer-Matrix-n-unit-cell}, the total transfer matrix can be written as
\begin{equation}
\widetilde{\bm{\Gamma}}^{n} = 
\cos(n\mu)\, \bm{I}_2 + \dfrac{\sin(n\mu)}{\sin\mu}
\begin{pmatrix}
0 & \widetilde{\Gamma}_{12} \\[3mm]
\widetilde{\Gamma}_{21} & 0
\end{pmatrix},
\label{mukka}
\end{equation}
in which $\bm{I}_2$ is the $2 \times 2$ identity matrix and $\mu$ is defined as 
\begin{equation}
\cos\mu = \frac{1}{2}\operatorname{tr}\widetilde{\bm{\Gamma}} = \widetilde{\Gamma}_{22}.
\label{eq:mu}
\end{equation}
For a cantilever rod, the boundary conditions are $\theta(0) = 0$ and $\theta'(L) = 0$, so that the second equation in the system \eqref{eq:y-transfered} becomes
\begin{equation}
\label{zio}
\cos(n \mu)\, \theta'(0) = \theta'(L) = 0.
\end{equation}
Since a nontrivial solution requires $\theta'(0) \neq 0$, Eq.~(\ref{zio}) leads to the condition
\begin{equation}
\cos(n \mu) = 0,
\end{equation}
which gives
\begin{equation}
\mu_m = \frac{(2m - 1) \pi}{2n}, \quad m = 1, 2, \ldots 
\end{equation}
The bifurcation condition is then obtained from Eq.~\eqref{eq:mu} as
\begin{equation}
\widetilde{\Gamma}_{22} - \cos\frac{(2m - 1) \pi}{2n} = 0,
\quad m = 1, 2, \ldots ,
\label{eq:bif-post-hom-micro-force}
\end{equation}
or equivalently as
\begin{equation}
\cos\Big(\beta_1(\ell_0 - 2\Delta)\Big) \cos\Big(2 \beta_2 \Delta\Big) - 
\frac{\beta_1^2 + \beta_2^2}{2\beta_1\beta_2}
\sin\Big(\beta_1(\ell_0 - 2\Delta)\Big) \sin\Big(2 \beta_2 \Delta\Big) - 
\cos \frac{(2m - 1)\pi}{2n} = 0.
\label{eq:bif-post-hom-axially-bendingly-model}
\end{equation}
To assess whether Model~(ii) reproduces the continuum limit of the discrete chain, let $n \to \infty$ while keeping the total length $L = n\ell_0$ and the internal length ratio $\delta = \Delta / \ell_0$ fixed. Consequently, $\ell_0 \to 0$ and $\Delta = \delta\, \ell_0 \to 0$.

Expanding the bifurcation condition \eqref{eq:bif-post-hom-axially-bendingly-model} to leading order in $\ell_0$, for fixed $m$, gives
\begin{equation}
\label{zia}
(1 - 2\delta) \beta_1^2 + 2 \delta \beta_2^2 = \frac{(2m - 1)^2 \pi^2}{4 L^2}.
\end{equation}
Substituting the definitions of $\beta_1^2$ and $\beta_2^2$ into Eq.~(\ref{zia}) and solving the resulting quadratic equation for $P$ yields
\begin{equation}
P_m^\pm = -\frac{1}{2} 
\left\{
EA + 4H\left(1 - 2\frac{\Delta}{\ell_0}\right) \pm \sqrt{EA^2 - (2m - 1)^2 \pi^2 \frac{EA\, EI}{L^2} - 32 H^2 \frac{\Delta}{\ell_0} \left(1 - 2\frac{\Delta}{\ell_0}\right)}
\right\},
\label{eq:limit2}
\end{equation}
for $m = 1, 2, \ldots$, provided that the discriminant is nonnegative.

Using the identifications $k_a \ell_0 = EA$ and $k_r \ell_0 = EI$, Eq.~\eqref{eq:limit2} can be compared directly with the critical loads obtained from the continuum limit of the original discrete chain, given in Eq.~\eqref{eq:limitchain}. The two results differ specifically in the terms involving the microforces $H$. Consequently, 
{\it 
retaining the microforces as periodically distributed external loads after homogenizing the elastic properties does not reproduce the correct continuum limit of the original discrete system}. 
This discrepancy shows that Model (ii) is not the appropriate homogenized model for the chain.

The origin of the discrepancy lies in the order in which the elastic properties and the microforces are homogenized. In the original discrete chain, the two outer portions of length $\Delta$ are inextensible, and the entire axial elongation of each cell occurs in the central spring, which carries the force $P+2H$. Model~(i) preserves this coupling by homogenizing the elastic energy and the work of the microforces simultaneously. 

In Model~(ii), by contrast, the axial stiffness is homogenized before the microforces are applied. The resulting rod is axially deformable throughout the cell. After the periodic microforces are introduced, Eq.~\eqref{eq:epsilon-post-hom-micro-force} gives the piecewise constant strain 
\begin{equation} 
\varepsilon^{(0)}(s) = 
\begin{cases} 
\dfrac{P+2H}{EA}, & \mathcal{I}(s)=1, \\[3mm] 
\dfrac{P}{EA}, & \mathcal{I}(s)=0, 
\end{cases} 
\end{equation} 
so that the central interval of relative length $1-2\delta$ carries the force $P + 2H$, whereas the two outer intervals, whose total relative length is $2\delta$, carry the force $P$. 

Using the definitions of $\beta_1^2$ and $\beta_2^2$, the effective geometric-stiffness coefficient obtained in the continuum limit of Model~(ii) is 
\begin{align} 
\mathcal{F}_{II}^{(0)} 
&= -EI\left[(1-2\delta)\beta_1^2+2\delta\beta_2^2\right] \nonumber \\ 
&= (1-2\delta)(P+2H) \left(1+\frac{P+2H}{EA}\right) + 2\delta P \left(1+\frac{P}{EA}\right). 
\label{eq:G-II-zero} 
\end{align}
By comparison, the corresponding coefficient previously obtained for Model~(i) is 
\begin{equation} 
\mathcal{F}_{I}^{(0)} = 
(P+2H) \left(1+\frac{P+2H}{EA}\right) - 4H\delta,  
\end{equation} 
hence the difference between the two coefficients is
\begin{equation} 
\label{bioa}
\mathcal{F}_{II}^{(0)} - \mathcal{F}_{I}^{(0)} = 
-\frac{8\delta H(P+H)}{EA}. 
\end{equation} 
The terms that are linear in $P$ and $H$ are therefore identical in the two models; the discrepancy arises from the nonlinear coupling between the local internal force and the axial stretch in Eq.~\eqref{eq:equil-post-hom-micro-force}. Consequently, homogenizing the elastic properties first and homogenizing the subsequently applied microforces afterward does not give the same result as homogenizing the loaded discrete chain directly. In other words, 
{\it 
these two homogenization operations do not commute}, 
explaining why Model~(ii) does not reproduce the correct continuum limit of the original discrete system.

To resolve the discrepancy represented by Eq.~(\ref{bioa}), Model~(iii) is introduced, in which only the flexural stiffness is homogenized, while the piecewise distribution of axial stiffness and the periodicity of the microforces are retained. In this way, the preservation of the microscale distribution of axial compliance will be shown to recover the correct continuum limit.

\subsection{Model (iii): Periodic microforces applied to a flexurally homogenized rod with piecewise axial stiffness}
\label{ssec:impose-forces-hom-flexurally-beam}

A third equivalent-rod model is introduced for the discrete chain described in Sec.~\ref{sec:1D-chain-discrete}. This model modifies Model (ii) proposed in Sec.~\ref{ssec:impose-forces-hom-axially-flexurally-beam} by homogenizing only the flexural stiffness while retaining a piecewise description of the axial stiffness:
\begin{equation}
EA(s) = 
\begin{cases}
(EA)_1, & (k-1)\ell_0 + \Delta < s < k\, \ell_0 - \Delta, \\[3mm]
(EA)_2, & \text{otherwise},
\end{cases}
\qquad
k=1,2,\ldots,n,
\label{eq:EA-piecewise}
\end{equation}
where $(EA)_1$ and $(EA)_2$ are the axial stiffnesses of the rod in the central interval and in the two outer intervals of the unit cell, respectively, as shown in Fig.~\ref{fig:post-hom-micro-force}. Note that, for now, the outer intervals are assumed to be axially deformable and not rigid as in the original model shown in Fig.~\ref{fig:1D-sketch}. 

The three parts of each unit cell act in series. Therefore, their combined axial stiffness must equal the stiffness $k_a$ of the axial spring in the discrete chain, which requires
\begin{equation}
\frac{1}{k_a}
=
\frac{\ell_0 - 2\Delta}{(EA)_1}
+
\frac{2\Delta}{(EA)_2}.
\label{eq:EA-calculated}
\end{equation}
Note that an axially homogeneous model is recovered by setting $(EA)_1=(EA)_2$. Indeed, Eq.~\eqref{eq:EA-calculated} gives in this case 
\begin{equation} 
(EA)_1=(EA)_2=k_a\ell_0=EA, 
\end{equation} 
consistently with Eq.~\eqref{eq:limit-kinematic-material2}.

Equation~\eqref{eq:EA-calculated} can alternatively be written as 
\begin{equation} 
(EA)_1 = k_a\ell_0 \left[ 1 - 2\frac{\Delta}{\ell_0} \left( 1-\frac{(EA)_1}{(EA)_2} \right) \right]. 
\label{eq:EA1-EA2} 
\end{equation} 
To reproduce the inextensible outer portions of the original discrete chain, the limit $(EA)_2\to\infty$ is considered while $(EA)_1$ remains finite. Eq.~\eqref{eq:EA1-EA2} then yields  
\begin{equation} 
(EA)_1 = k_a\ell_0 \left( 1-2\frac{\Delta}{\ell_0} \right), 
\label{eq:EA1} 
\end{equation} 
which, introducing $\delta=\Delta/\ell_0$ and using $EA=k_a\ell_0$, becomes 
\begin{equation} 
(EA)_1=EA(1-2\delta). 
\label{eq:EA1-delta} 
\end{equation}
The rotational equilibrium equation has the same form as in Model~(ii), Eq.~\eqref{eq:equil-post-hom-micro-force}:
\begin{equation}
\theta''(s) - \frac{P + 2H\, \mathcal{I}(s)}{EI}
(1+\varepsilon(s))\sin\theta(s) + \frac{V}{EI} (1+\varepsilon(s)) \cos\theta(s) = 0, 
\label{eq:equil-post-hom-micro-force-app}
\end{equation}
however, because the axial stiffness is piecewise constant, the axial strain is now
\begin{equation}
\varepsilon(s)=
\begin{cases}
\dfrac{(P+2H)\cos\theta(s) + V\sin\theta(s)}{(EA)_1},
& (k-1)\ell_0 + \Delta < s < k\, \ell_0 - \Delta, \\[3mm]
\dfrac{P\cos\theta(s) + V\sin\theta(s)}{(EA)_2},
& \text{otherwise},
\end{cases}
\qquad
k=1,2,\ldots,n.
\label{eq:epsilon-post-hom-micro-force-app}
\end{equation}
In the limit $(EA)_2 \to \infty$, the two outer parts of the chain remain axially unstrained, consistently with the original kinematics, Fig.~\ref{fig:1D-sketch}. 

To determine the bifurcation loads for a cantilever rod, $V = 0$ is set and the equilibrium and strain equations are linearized about the straight configuration. The resulting equation has the same form as that obtained for Model (ii), Eq.~\eqref{eq:equilibrium-post-hom-micro-force}:
\begin{equation}
\theta''(s) + \beta^2(s) \theta(s) = 0,
\label{eq:modeliii}
\end{equation}
but, differently from Model (ii), $\beta^2(s)$ is now replaced by
\begin{equation}
\beta^2(s)=
\begin{cases}
\beta_1^2 = - \dfrac{P + 2H}{EI} \left(1 + \dfrac{P + 2H}{(EA)_1}\right),
& (k-1)\ell_0 + \Delta < s < k\ell_0 - \Delta, 
\\[5mm]
\beta_2^2 = - \dfrac{P}{EI} \left( 1 + \dfrac{P}{(EA)_2} \right),
& \text{otherwise},
\end{cases}
\qquad k=1,2,\ldots,n.
\label{eq:newbeta}
\end{equation}
For $(EA)_2\to\infty$, the second coefficient reduces to 
\begin{equation} 
\beta_2^2 = -\frac{P}{EI}. 
\end{equation} 
The bifurcation condition is therefore the same condition obtained for Model (ii), but with $\beta_1$ and $\beta_2$ defined by Eq.~\eqref{eq:newbeta}.

To determine the continuum limit, let $\ell_0\to0$ and $n=L/\ell_0\to\infty$, while keeping $\delta=\Delta/\ell_0$ fixed. As in Model~(ii), an expansion of the bifurcation condition to leading order in $\ell_0$, for fixed $m$, yields  
\begin{equation} 
(1-2\delta)\beta_1^2 + 2\delta\beta_2^2 = \frac{(2m-1)^2\pi^2}{4L^2}, 
\qquad m=1,2,\ldots 
\label{eq:model-iii-continuum-condition} 
\end{equation} 
The effective geometric-stiffness parameter for Model~(iii) is therefore 
\begin{equation} 
\mathcal{F}_{III}^{(0)} = -EI \left[ (1-2\delta)\beta_1^2 + 2\delta\beta_2^2 \right],  
\label{eq:G-III-zero} 
\end{equation} 
which, using $(EA)_1=EA(1-2\delta)$ and $(EA)_2\to\infty$, leads to  
\begin{equation}
\mathcal{F}_{III}^{(0)} 
= (P+2H) \left( 1+\frac{P+2H}{EA} \right) - 4H\delta. 
\label{eq:G-III-expanded}    
\end{equation} 
The right-hand side of the above equation is precisely the effective geometric-stiffness parameter obtained for Model~(i) and hence 
\begin{equation} 
\mathcal{F}_{III}^{(0)} = \mathcal{F}_{I}^{(0)}. 
\label{eq:G-III-G-I} 
\end{equation} 
Using the identifications $EA=k_a\ell_0$ and $EI=k_r\ell_0$, Eq.~\eqref{eq:model-iii-continuum-condition} yields the same continuum-limit bifurcation loads as Model~(i) and the original discrete chain. Thus, retaining the piecewise distribution of axial stiffness restores the correct coupling between the microforces and the axial compliance that is lost in Model~(ii). 

This agreement concerns the bifurcation condition in the continuum limit. Unlike Model~(i), Model~(iii) still retains the microforces as explicitly applied periodic external loads and therefore does not provide the same fully homogenized nonlinear formulation.

\section{Closed-form solution for the large deflections of the homogenized rod Model (i)}
\label{app:solutions-elliptic}

Homogenization serves several purposes in the analysis of discrete systems. It replaces a structure with many elements, $n\gg1$, by an equivalent continuum, avoiding the need to determine every microscopic unknown. It can also clarify the mechanical interpretation of continuum theories, as in the distinction between the Engesser and Haringx--Reissner rod models~\cite{Kocsis} and in the transformation of microscale loading into macroscopic contributions identified in the previous section. A further advantage is the possibility of obtaining closed-form solutions. The large-deflection solution derived below provides one example; the weakly nonlinear analysis in Appendix~\ref{app:weakly-nonlinear-clamped-beam} provides another.

Following~\cite{humer_buckling_2013,batista2016closed}, an analytical solution is derived for the homogenized rod in Model (i), governed by Eq.~\eqref{eq:equi-hom-sim}. Consider a cantilever rod, subjected only to an axial end force, so that $V = 0$. Introducing the dimensionless coordinate $\xi = s/L$, Eq.~\eqref{eq:equi-hom-sim} becomes
\begin{equation}
\kappa(\xi)=\theta'(\xi),
\qquad
\theta''(\xi) 
- \frac{1}{\chi} \left[(p+2h)+(p+2h)^2\cos\theta(\xi) - 4h\delta\right]
\sin\theta(\xi)=0.
\label{eq:equi-hom-sim-for-elliptic}
\end{equation}
Here and below, primes denote differentiation with respect to $\xi$, and $\kappa = d\theta/d\xi = Ld\theta/ds$ is the dimensionless curvature. The displacement components $u$ and $v$ are made dimensionless through division by $L$. The boundary conditions at the clamp and at the loaded end are
\begin{equation}
\theta(0)=0,
\qquad
\kappa(1)=\theta'(1)=0, 
\label{eq:1D-hom-horizontal-load-bc-dimensionless}
\end{equation}
respectively. Let $\theta_1=\theta(1)$ denote the free-end rotation; the corresponding curvature is $\kappa_1=\kappa(1)=0$. Multiplying Eq.~\eqref{eq:equi-hom-sim-for-elliptic}$_2$ by $\theta'$ gives
\begin{equation}
\frac{d}{d\xi}
\left\{
\frac{\kappa^2}{2}
-\frac{4h\delta-(p+2h)}{\chi}\cos\theta
+\frac{(p+2h)^2}{2\chi}\cos^2\theta
\right\}=0.
\label{eq:first-integral-derivative}
\end{equation}
Evaluating the integration constant at $\xi=1$ yields
\begin{equation}
\kappa^2
=
\frac{2[4h\delta-(p+2h)]}{\chi}
(\cos\theta-\cos\theta_1)
-\frac{(p+2h)^2}{\chi}
(\cos^2\theta-\cos^2\theta_1).
\label{eq:first-integral-dimensional-parameters}
\end{equation}
The governing equation and its boundary conditions are invariant under $\theta\mapsto-\theta$ and $\kappa\mapsto-\kappa$. It is therefore sufficient to consider the representative branch with positive tip rotation and define
\begin{equation}
k=\sin\frac{\theta_1}{2},
\qquad
0<k<1.
\label{eq:elliptic-amplitude}
\end{equation}
A reflected equilibrium configuration can simply be obtained by changing the signs of $\theta$ and $\kappa$. The value $k=0$ corresponds to the trivial straight solution, whereas $k=1$ corresponds to $\theta_1=\pi$. The latter can be approached but cannot be reached by a rod of finite length subject to a finite load.

Evaluating Eq.~\eqref{eq:first-integral-dimensional-parameters} at the clamped end gives
\begin{equation}
\kappa^2(0)
=
\frac{4k^2}{\chi}
\left\{
4h\delta-(p+2h)-(p+2h)^2(1-k^2)
\right\}, 
\label{eq:clamped-curvature-admissibility}
\end{equation}
so that a real nontrivial equilibrium must  satisfy the following ellipticity condition
\begin{equation}
4h\delta-(p+2h)-(p+2h)^2(1-k^2)>0, 
\label{eq:dimensional-elliptic-admissibility}
\end{equation}
where equality is excluded. The latter statement can be proved as follows. Equality in the ellipticity condition would imply $\kappa(0)=0$, which, together with $\theta(0)=0$, provides zero initial data for an initial-value problem governed by a first-order system with a smooth right-hand side. The uniqueness theorem for ordinary differential equations therefore implies that the solution is identically zero, $\theta(\xi)=\kappa(\xi)=0$~\cite[Sec.~2.2, Theorem~2.2]{teschl_ordinary_2012}. Thus, equality is possible only for the trivial equilibrium, and the inequality must be strict for every nontrivial equilibrium.

Note that Eq.~(\ref{eq:dimensional-elliptic-admissibility}) implies $4h\delta-(p+2h) > 0$, which motivates the following notation  
\begin{equation}
\Omega^2
=
\frac{4h\delta-(p+2h)}{\chi},
\qquad
\nu
=
\frac{(p+2h)^2}{4h\delta-(p+2h)}, 
\label{eq:omega-mu}
\end{equation}
so that the admissibility condition~\eqref{eq:dimensional-elliptic-admissibility} becomes
\begin{equation}
1-\nu(1-k^2)>0, 
\label{eq:elliptic-reality-condition}
\end{equation}
and the admissible interval for $k$ is therefore
\begin{equation}
\begin{cases}
0<k<1, & 0\leq\nu\leq1, \\[2mm]
\displaystyle\sqrt{1-\frac{1}{\nu}}<k<1, & \nu>1.
\end{cases}
\label{eq:admissible-k-interval}
\end{equation}
The transcendental equation derived below must be solved within the above-defined interval.

Using Eq.~\eqref{eq:omega-mu}, the first integral assumes the form
\begin{equation}
\kappa^2
=
4\Omega^2
\left(
k^2-\sin^2\frac{\theta}{2}
\right)
\left[
1-\nu
+\nu\left(
k^2+\sin^2\frac{\theta}{2}
\right)
\right].
\label{eq:kappaSauqre-1}
\end{equation}
Introducing the change of variable 
\begin{equation}
\sin\frac{\theta}{2}=k\sin\phi,
\label{eq:elliptic-variable}
\end{equation}
Only the principal branch $|\theta|<\pi$ is analyzed. Under the transformation~(\ref{eq:elliptic-variable}), Eq.~\eqref{eq:kappaSauqre-1} becomes
\begin{equation}
\kappa^2
=
4\Omega^2k^2\cos^2\phi
\left[1-\nu+\nu k^2(1+\sin^2\phi)\right] ,
\label{eq:kappaSauqre-2}
\end{equation}
while a differentiation of Eq.~\eqref{eq:elliptic-variable} leads to
\begin{equation}
\frac{1}{2}\cos\frac{\theta}{2}\frac{d\theta}{d\xi}
=
k\cos\phi\frac{d\phi}{d\xi},
\end{equation}
so that the following equation is obtained
\begin{equation}
\kappa
=
\frac{2k\cos\phi}{\sqrt{1-k^2\sin^2\phi}}
\frac{d\phi}{d\xi}.
\label{eq:kappa}
\end{equation}
A substitution of the above equation into Eq.~\eqref{eq:kappaSauqre-2} yields
\begin{equation}
\left(\frac{d\phi}{d\xi}\right)^2
=
\Omega^2(1-k^2\sin^2\phi)
\left[1-\nu+\nu k^2(1+\sin^2\phi)\right].
\label{eq:f-deriv-Square}
\end{equation}
For $k$ in the interval~\eqref{eq:admissible-k-interval}, the definitions 
\begin{equation}
\tilde{\Omega}^2
=
\Omega^2\left[1+\nu(2k^2-1)\right],
\qquad
\eta^2
=
\frac{\nu k^2}{1-\nu(1-k^2)} ,
\label{eq:omegaTilde-m}
\end{equation}
are introduced and the nonnegative roots $\eta\geq0$ and $\tilde{\Omega}>0$ are selected.

Equation~\eqref{eq:f-deriv-Square} can be rewritten as
\begin{equation}
\left(\frac{d\phi}{d\xi}\right)^2
=
\frac{\tilde{\Omega}^2}{1+\eta^2}
(1-k^2\sin^2\phi)(1+\eta^2\sin^2\phi).
\label{eq:f-deriv}
\end{equation}
Setting
\begin{equation}
\psi \equiv \sin\phi ,
\label{eq:mu-defined}
\end{equation}
transforms Eq.~\eqref{eq:f-deriv} into
\begin{equation}
\left(\frac{d\psi}{d\xi}\right)^2
=
\frac{\tilde{\Omega}^2}{1+\eta^2}
(1-\psi^2)(1-k^2\psi^2)(1+\eta^2\psi^2).
\label{eq:mu-equation}
\end{equation}
By defining a new variable $t$ through the Gaussian transformation
\begin{equation}
\psi = \frac{t}{\sqrt{1 + \eta^2(1 - t^2)}}.
\label{eq:Gaussian-transfor}
\end{equation}
Equation~\eqref{eq:mu-equation} reduces to
\begin{equation}
\left(\frac{dt}{d\xi}\right)^2
=
\tilde{\Omega}^2(1-t^2)(1-\tilde{k}^2t^2),
\qquad
\tilde{k}^2=\frac{k^2+\eta^2}{1+\eta^2}.
\label{eq:t-equation}
\end{equation}
The admissibility condition and $0<k<1$ ensure that $0<\tilde{k}<1$. A representative real solution of Eq.~\eqref{eq:t-equation} is
\begin{equation}
t(\xi)=\operatorname{sn}(\tilde{\Omega}\xi+C,\tilde{k}),
\label{eq:t}
\end{equation}
where $\operatorname{sn}(u,k)=\sin(\operatorname{am}(u,k))$ is the Jacobi elliptic sine, $\operatorname{am}(\cdot,\cdot)$ is the Jacobi amplitude, and $C$ is an integration constant. A back-substitution into Eqs.~\eqref{eq:Gaussian-transfor}, \eqref{eq:mu-defined}, and~\eqref{eq:elliptic-variable}, followed by differentiation with respect to $\xi$, gives
\begin{equation}
\theta(\xi)
=
2\arcsin
\left[
\frac{k\operatorname{sn}(\tilde{\Omega}\xi+C,\tilde{k})}
{\sqrt{1+\eta^2\operatorname{cn}^2(\tilde{\Omega}\xi+C,\tilde{k})}}
\right],
\qquad
\kappa(\xi)
=
\frac{2\tilde{\Omega}k\sqrt{1+\eta^2}
\operatorname{cn}(\tilde{\Omega}\xi+C,\tilde{k})}
{1+\eta^2\operatorname{cn}^2(\tilde{\Omega}\xi+C,\tilde{k})} ,
\label{sol-rotation-curvature}
\end{equation}
where, $\operatorname{cn}(u,k)=\cos(\operatorname{am}(u,k))$ is the Jacobi elliptic cosine. 

The free-end condition $\kappa(1)=0$ gives
\begin{equation}
\operatorname{cn}(\tilde{\Omega}+C,\tilde{k})=0,
\end{equation}
which implies
\begin{equation}
C=(2r-1)K(\tilde{k})-\tilde{\Omega},
\qquad
r\in\mathbb{Z},
\label{eq:C-sol-1}
\end{equation}
where $K(\cdot)$ is the complete elliptic integral of the first kind. The condition at the clamp, $\theta(0)=0$, can be written as
\begin{equation}
\operatorname{sn}(C,\tilde{k})=0,
\end{equation}
and hence
\begin{equation}
C=2jK(\tilde{k}),
\qquad
j\in\mathbb{Z}.
\label{eq:C-sol-2}
\end{equation}
Equating Eqs.~\eqref{eq:C-sol-1} and~\eqref{eq:C-sol-2} leads to the transcendental equation
\begin{equation}
\tilde{\Omega}-(2q-1)K(\tilde{k})=0,
\qquad
q=r-j=1,2,\ldots ,
\label{eq:elliptic-transcendental}
\end{equation}
where, $q$ indexes the equilibrium branches. For the representative solution, $\theta_1>0$, it is set $r=1$, or equivalently $j=1-q$, to obtain
\begin{equation}
C=2(1-q)K(\tilde{k}).
\label{eq:elliptic-phase-positive-tip}
\end{equation}
With this choice, $\tilde{\Omega}+C=K(\tilde{k})$. Since $\operatorname{cn}(K(\tilde{k}),\tilde{k})=0$ and $\operatorname{sn}(K(\tilde{k}),\tilde{k})=1$, the free-end curvature vanishes and $\theta_1=2\arcsin k>0$. For prescribed load parameters $p$ and $h$, Eq.~\eqref{eq:elliptic-transcendental} must generally be solved numerically over the interval~\eqref{eq:admissible-k-interval}. If roots are not found within this interval, the corresponding branch contains no real nontrivial equilibrium. Once $k$ is known, the axial strain follows from
\begin{equation}
\varepsilon = (p + 2 h)\cos\theta = (p + 2 h) \frac{1 - 2 k^2 + (\eta^2 + 2 k^2)\operatorname{cn}^2(\tilde{\Omega} \xi + C,\tilde{k})}{1 + \eta^2 \operatorname{cn}^2(\tilde{\Omega} \xi + C,\tilde{k})} .
\label{eq:strain-solved}
\end{equation}
In analogy to Eq.~\eqref{eq:condition-admissibility}, a mechanically admissible configuration requires the current length of the central axial spring in the original discrete chain to remain positive, so that
\begin{equation}
\lambda(\xi)
=
1+\varepsilon(\xi)
=
1+(p+2h)\cos\theta(\xi)>2\,\delta,
\qquad
0\leq\xi\leq1.
\label{eq:positive-stretch-elliptic}
\end{equation}
This requirement is distinct from the ellipticity admissibility condition~\eqref{eq:elliptic-reality-condition}. Since $1-2k^2 \leq \cos\theta \leq 1$ on the branch considered here, Eq.~(\ref{eq:positive-stretch-elliptic}) is equivalent to 
\begin{equation}
\begin{cases}
1+(p+2h)(1-2k^2)>2\,\delta, & p+2h\geq0, \\[2mm]
1+p+2h>2\,\delta, & p+2h<0.
\end{cases}
\label{eq:positive-stretch-elliptic-explicit}
\end{equation}
Substituting Eqs.~\eqref{sol-rotation-curvature} and \eqref{eq:strain-solved} into the dimensionless kinematic relations provides the derivatives of the displacement components
\begin{equation}
u' = (1 + (p + 2 h)\cos\theta)\cos\theta - 1, \qquad\qquad v' = (1 + (p + 2 h)\cos\theta)\sin\theta ,
\label{eq:uprime-vprime_2}
\end{equation}
so that $u$ and $v$ follow by integration. The rigid-body translation is fixed by imposing the clamp constraint at the origin:
\begin{equation}
u(0)=0,
\qquad
v(0)=0.
\label{eq:clamped-translations-elliptic}
\end{equation}
For the integrations, the variable 
\begin{equation}
z = \tilde{\Omega} \xi + C,
\label{z}
\end{equation}
is introduced, so that $z=C$ at $\xi=0$. The following formulas assume $\eta>0$, while the regular limit $\eta=0$ is given separately below. Adopting the shorthand notation 
\begin{equation}
\operatorname{sn} \equiv \operatorname{sn}(z,\tilde{k}), \qquad\qquad \operatorname{cn} \equiv \operatorname{cn}(z,\tilde{k}), \qquad\qquad \operatorname{dn} \equiv \operatorname{dn}(z,\tilde{k}),
\end{equation}
where $\operatorname{dn}^2 = 1 - \tilde{k}^2 \operatorname{sn}^2$, the derivatives with respect to $z$ satisfy
\begin{equation}
\operatorname{sn}' = \operatorname{cn} \operatorname{dn}, \qquad\qquad \operatorname{cn}' = -\operatorname{sn} \operatorname{dn}, \qquad\qquad \operatorname{dn}' = -\tilde{k}^2 \operatorname{sn} \operatorname{cn} .
\end{equation}
By the chain rule,
\[
\frac{dv}{d\xi} = \frac{dv}{d\operatorname{cn}} \frac{d\operatorname{cn}}{d\xi} = -\tilde{\Omega} \operatorname{sn} \operatorname{dn} \frac{dv}{d\operatorname{cn}},
\]
and Eq.~\eqref{eq:uprime-vprime_2}$_2$ becomes
\begin{equation}
\label{marianna}
\begin{aligned}
\frac{dv}{d\operatorname{cn}}
&=
-\frac{2k\sqrt{1+\eta^2}}{\tilde{\Omega}}
\Biggl\{
\frac{1}{1+\eta^2\operatorname{cn}^2}
+
(p+2h)
\frac{1-2k^2+(\eta^2+2k^2)\operatorname{cn}^2}{(1+\eta^2\operatorname{cn}^2)^2}
\Biggr\}
\\[0.8em]
&=
-\frac{2k\sqrt{1+\eta^2}}{\tilde{\Omega}}
\begin{array}[t]{@{}l@{}}
\displaystyle
\Biggl\{
\frac{1}{1+\eta^2\operatorname{cn}^2}
+
(p+2h)
\Biggl(
-\frac{k^2(1+\eta^2)}{\eta^2}
\frac{1-\eta^2\operatorname{cn}^2}{(1+\eta^2\operatorname{cn}^2)^2}
\\[0.25em]
\multicolumn{1}{r@{}}{\displaystyle
+
\left(
1-k^2+\frac{k^2}{\eta^2}
\right)
\frac{1}{1+\eta^2\operatorname{cn}^2}
\Biggr)
\Biggr\}}
\end{array}
\\[0.8em]
&=
-\frac{2k\sqrt{1+\eta^2}}{\tilde{\Omega}}
\begin{array}[t]{@{}l@{}}
\displaystyle
\Biggl\{
-(p+2h)
\frac{k^2(1+\eta^2)}{\eta^2}
\frac{1-\eta^2\operatorname{cn}^2}
     {(1+\eta^2\operatorname{cn}^2)^2}
\\[0.25em]
\multicolumn{1}{r@{}}{\displaystyle
+
\left(
1+(p+2h)
\left(
1-k^2+\frac{k^2}{\eta^2}
\right)
\right)
\frac{1}{1+\eta^2\operatorname{cn}^2}
\Biggr\}}.
\end{array}
\end{aligned}
\end{equation}
Integrating Eq.~\eqref{marianna} and imposing $v(z=C) = v(\xi=0) = 0$ provides the transverse displacement
\begin{equation}
\begin{aligned}
v(z) = \frac{2 k \sqrt{1 + \eta^2}}{\tilde{\Omega}} \Biggl\{
&\frac{1 + (p + 2 h)\left( 1 - k^2 + \frac{k^2}{\eta^2} \right)}{\eta}
\left( \arctan \bigl(\eta \operatorname{cn}(C,\tilde{k})\bigr) - \arctan \bigl(\eta \operatorname{cn}(z,\tilde{k})\bigr) \right) \\[3mm]
&\left. - (p + 2 h) \frac{k^2(1 + \eta^2)}{\eta^2}
\left( \frac{\operatorname{cn}(C,\tilde{k})}{1 + \eta^2 \operatorname{cn}^2(C,\tilde{k})} - \frac{\operatorname{cn}(z,\tilde{k})}{1 + \eta^2 \operatorname{cn}^2(z,\tilde{k})} \right) \right\}.
\end{aligned}
\label{eq:v-final}
\end{equation}
The integration of Eq.~\eqref{eq:uprime-vprime_2}$_1$ is facilitated by the notation
\begin{equation}
\cos\theta
= 1 - \frac{2 k^2 \operatorname{sn}^2}{1 + \eta^2 \operatorname{cn}^2}
= \mathcal{S}_1 - \frac{\mathcal{S}_2}{1 + \eta^2 \operatorname{cn}^2} ,
\end{equation}
where
\begin{equation}
\mathcal{S}_1 = 1 + \frac{2 k^2}{\eta^2}, \qquad\qquad \mathcal{S}_2 = \frac{2 k^2 (1 + \eta^2)}{\eta^2} .
\end{equation}
It follows that
\begin{equation}
\begin{aligned}
u'
&= \cos\theta - 1 + (p + 2 h)\cos^2\theta \\[3mm]
&= \left( \mathcal{S}_1 - \frac{\mathcal{S}_2}{1 + \eta^2 \operatorname{cn}^2} \right) - 1
+ (p + 2 h) \left( \mathcal{S}_1 - \frac{\mathcal{S}_2}{1 + \eta^2 \operatorname{cn}^2} \right)^2 \\[3mm]
&= (p + 2 h) \mathcal{S}_1^2 + \mathcal{S}_1 - 1
- \frac{\bigl( 2 (p + 2 h) \mathcal{S}_1 + 1 \bigr) \mathcal{S}_2}{1 + \eta^2 \operatorname{cn}^2}
+ \frac{(p + 2 h) \mathcal{S}_2^2}{(1 + \eta^2 \operatorname{cn}^2)^2} .
\end{aligned}
\end{equation}
Integrating from $C$ to $z$ leads to a useful expression for the displacement component $u$
\begin{equation}
\begin{aligned}
u
&=
\frac{(p + 2h) \mathcal{S}_1^2 + \mathcal{S}_1 - 1}{\tilde{\Omega}} (z - C)
\\[3mm]
&\quad
-\frac{\bigl(2 (p + 2h) \mathcal{S}_1 + 1\bigr) \mathcal{S}_2}{\tilde{\Omega}}
 \int_C^z \frac{d\zeta}{1 + \eta^2 \operatorname{cn}^2(\zeta,\tilde{k})}
+\frac{(p + 2h) \mathcal{S}_2^2}{\tilde{\Omega}}
 \int_C^z \frac{d\zeta}{\bigl(1 + \eta^2 \operatorname{cn}^2(\zeta,\tilde{k})\bigr)^2}.
\end{aligned}
\end{equation}
The first integral is
\begin{equation}
\int_C^z \frac{d\zeta}{1 + \eta^2 \operatorname{cn}^2(\zeta,\tilde{k})}
= \frac{1}{1 + \eta^2} \int_C^z \frac{d\zeta}{1 - \rho \operatorname{sn}^2(\zeta,\tilde{k})}
= \frac{1}{1 + \eta^2} \left[ \Pi(\operatorname{am}(z,\tilde{k}), \rho, \tilde{k}) - \Pi(\operatorname{am}(C,\tilde{k}), \rho, \tilde{k}) \right],
\end{equation}
where
\begin{equation}
\rho = \frac{\eta^2}{1 + \eta^2} .
\end{equation}
To evaluate the second integral, the following antiderivative is used
\begin{equation}
\begin{aligned}
\int \frac{d z}{(1 + \eta^2 \operatorname{cn}^2)^2}
&= \frac{1}{(1 + \eta^2)^2} \int \frac{d z}{(1 - \rho \operatorname{sn}^2)^2} \\[3mm]
&= \frac{1}{(1 + \eta^2)^2} \Biggl( a_1 z + a_2 E(\operatorname{am}(z,\tilde{k}), \tilde{k}) + a_3 \Pi(\operatorname{am}(z,\tilde{k}), \rho, \tilde{k}) \\[3mm]
&\qquad\qquad\qquad\quad + a_4 \frac{\operatorname{sn}(z,\tilde{k}) \operatorname{cn}(z,\tilde{k}) \operatorname{dn}(z,\tilde{k})}{1 - \rho \operatorname{sn}^2(z,\tilde{k})} \Biggr),
\end{aligned}
\end{equation}
where $a_1,\ldots,a_4$ are constants to be determined, $E$ and $\Pi$ denote the incomplete elliptic integrals of the second and third kinds, respectively, and are defined as 
\begin{equation}
E(\phi,k) \equiv \int_0^{\phi} \sqrt{1 - k^2 \sin^2\vartheta} d\vartheta,
\qquad
\Pi(\phi,\rho,k) \equiv \int_0^{\phi} \frac{1}{(1 - \rho \sin^2\vartheta)\sqrt{1 - k^2 \sin^2\vartheta}} d\vartheta .
\end{equation}
For higher branches, the Jacobi amplitude and the associated incomplete elliptic integrals are understood as continuous functions along the real $z$-axis. This convention avoids artificial jumps at period boundaries.

Using the identities 
\begin{equation}
\frac{d}{d z} E(\operatorname{am}(z,\tilde{k}), \tilde{k}) = \operatorname{dn}^2 = 1 - \tilde{k}^2 \operatorname{sn}^2,
\qquad
\frac{d}{d z} \Pi(\operatorname{am}(z,\tilde{k}), \rho, \tilde{k}) = \frac{1}{1 - \rho \operatorname{sn}^2},
\end{equation}
and
\begin{equation}
\left( \frac{\operatorname{sn} \operatorname{cn} \operatorname{dn}}{1 - \rho \operatorname{sn}^2} \right)'
= \frac{(\operatorname{sn} \operatorname{cn} \operatorname{dn})'(1 - \rho \operatorname{sn}^2) + 2 \rho \operatorname{sn}^2 \operatorname{cn}^2 \operatorname{dn}^2}{(1 - \rho \operatorname{sn}^2)^2},
\end{equation}
allows the explicit determination of the four constants $a_1,\ldots,a_4$  
\begin{equation}
a_1 = -\frac{1 + \eta^2}{2}, \quad
a_2 = -\frac{\eta^2 (1 + \eta^2)}{2 k^2}, \quad
a_3 = \frac{\eta^2 k^2 + \eta^2 + 3 k^2}{2 k^2}, \quad
a_4 = \frac{\eta^4}{2 k^2} .
\end{equation}
Therefore,
\begin{equation}
\int_C^z
\frac{d\zeta}{\bigl(1+\eta^2\operatorname{cn}^2(\zeta,\tilde{k})\bigr)^2}
=
\mathcal{J}(z)-\mathcal{J}(C),
\end{equation}
where
\begin{multline}
\mathcal{J}(z)
= -\frac{z}{2 (1 + \eta^2)}
- \frac{\eta^2}{2 k^2 (1 + \eta^2)} E(\operatorname{am}(z,\tilde{k}), \tilde{k})
+ \frac{\eta^2 k^2 + \eta^2 + 3 k^2}{2 k^2 (1 + \eta^2)^2}
\Pi(\operatorname{am}(z,\tilde{k}), \rho, \tilde{k})
\\[3mm]
+ \frac{\eta^4}{2 k^2 (1 + \eta^2)}
\frac{\operatorname{sn}(z,\tilde{k}) \operatorname{cn}(z,\tilde{k}) \operatorname{dn}(z,\tilde{k})}
{1 + \eta^2 \operatorname{cn}^2(z,\tilde{k})}.
\end{multline}
The longitudinal displacement assumes the form
\begin{equation}
\begin{aligned}
u(z)
&= \frac{(p + 2 h) \mathcal{S}_1^2 + \mathcal{S}_1 - 1}{\tilde{\Omega}} (z - C)
+ \frac{(p + 2 h) \mathcal{S}_2^2}{\tilde{\Omega}} \left( \mathcal{J}(z) - \mathcal{J}(C) \right) \\[3mm]
&\qquad + \frac{-\bigl( 2 (p + 2 h) \mathcal{S}_1 + 1 \bigr) \mathcal{S}_2}{\tilde{\Omega} (1 + \eta^2)}
\left[ \Pi(\operatorname{am}(z,\tilde{k}), \rho, \tilde{k}) - \Pi(\operatorname{am}(C,\tilde{k}), \rho, \tilde{k}) \right] .
\end{aligned}
\label{eq:u-final}
\end{equation}
In the nontrivial limiting case $\eta=0$ with $k>0$, which occurs when the rod is axially unstretched, $p+2h=0$, the apparent singularities in Eqs.~\eqref{eq:v-final} and \eqref{eq:u-final} are removable. In this case, $\tilde{\Omega}=\Omega$, $\tilde{k}=k$, and $\varepsilon=0$. The displacement components satisfying Eq.~\eqref{eq:clamped-translations-elliptic} reduce to
\begin{equation}
v(z) = \frac{2k}{\Omega}
\left[
\operatorname{cn}(C,k)-\operatorname{cn}(z,k)
\right],
\label{eq:v-final-m-zero}
\end{equation}
and
\begin{equation}
u(z)
=
-\frac{2}{\Omega}
\left[
(z-C)-E(\operatorname{am}(z,k),k)+E(\operatorname{am}(C,k),k)
\right].
\label{eq:u-final-m-zero}
\end{equation}
Equations~\eqref{eq:u-final} and~\eqref{eq:v-final}, together with the limiting expressions~\eqref{eq:v-final-m-zero} and \eqref{eq:u-final-m-zero}, complete the analytical solution for the homogenized cantilever rod subjected to a horizontal end force. Comparison with numerical solutions obtained by a shooting method in Wolfram Mathematica (version 15.0), which are not reported for brevity, showed agreement for all parameter values investigated.

\section{Comparison between the different models for a cantilever rod}
\label{ssec:comparison-bifurcation}

\subsection{Bifurcation of the cantilever rod}

\begin{figure}[H]
    \centering
    \includegraphics[width=1\linewidth]{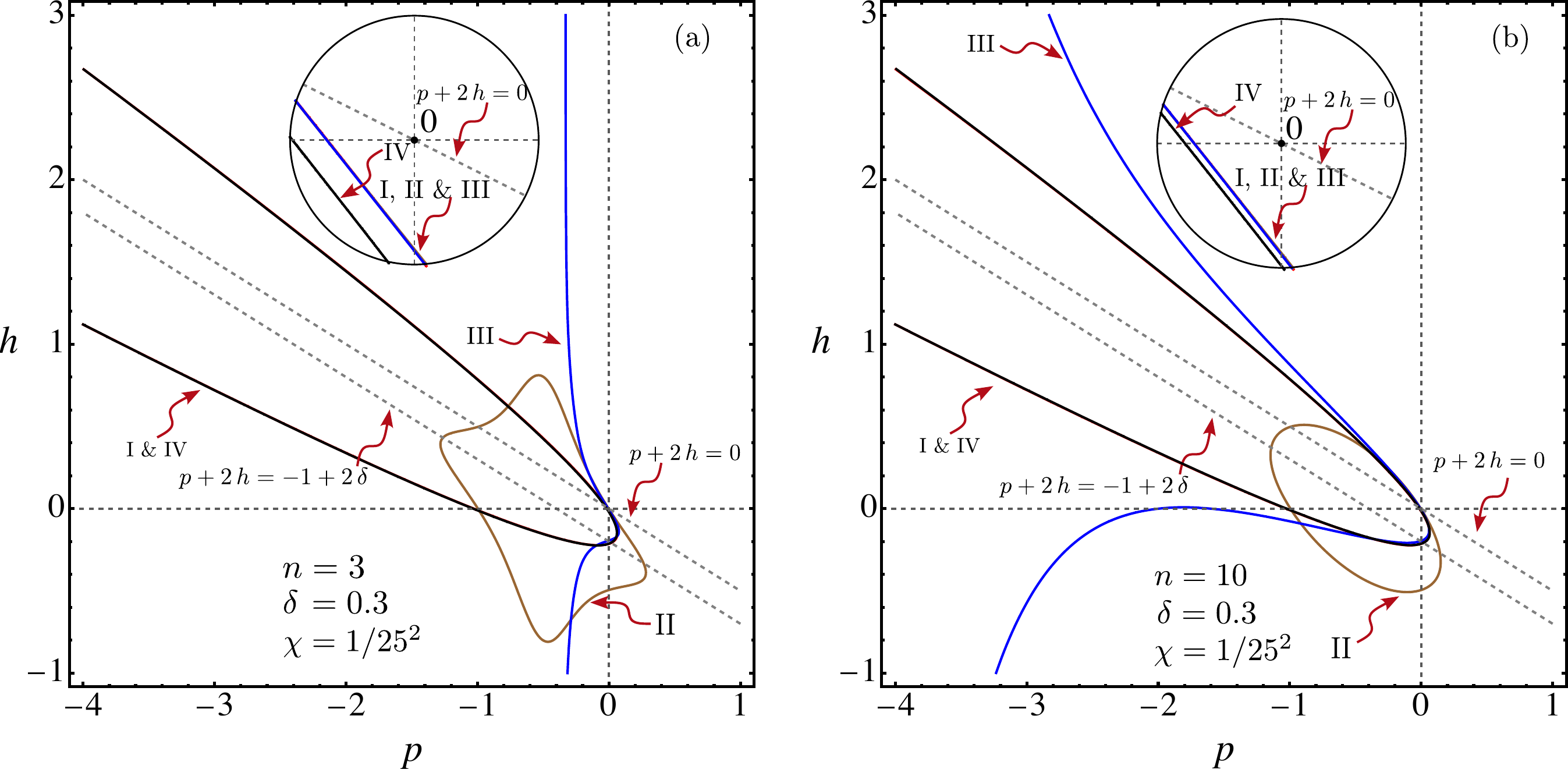}
    \caption{
    Critical combinations of $p$ and $h$ at bifurcation as predicted by the three homogenized rod Models~(i)--(iii), and the corresponding discrete chain Model~(iv), for $\chi = 1/25^2$ and $\delta = 0.3$: (a) $n = 3$ and (b) $n = 10$. Model~(i) remains highly accurate even for very small numbers of elements, although homogenization is generally expected to be most reliable for $n \gg 1$.
    }
    \label{fig:bif-a}
\end{figure}

\begin{figure}[H]
    \centering
    \includegraphics[width=1\linewidth]{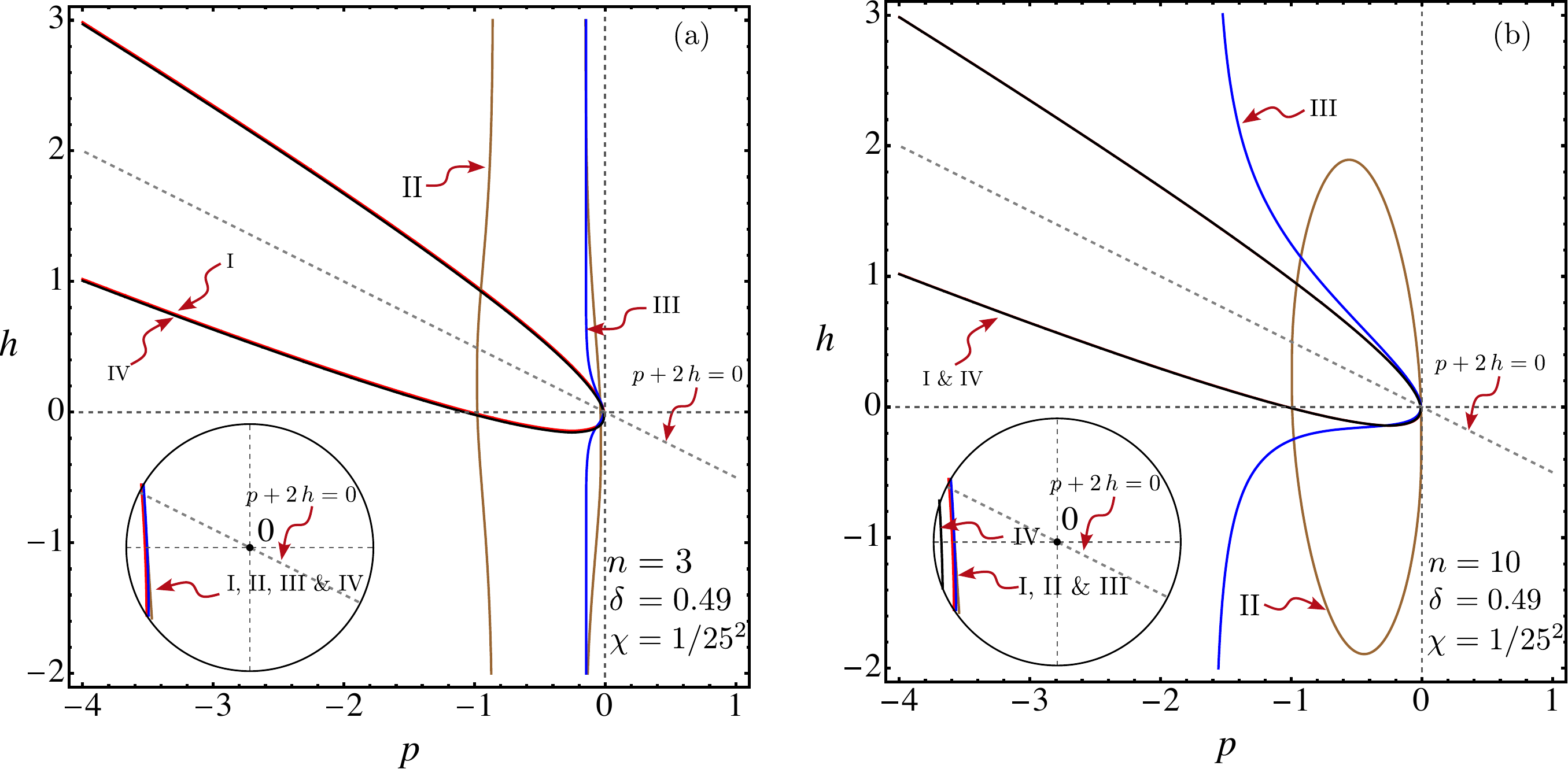}
    \caption{
    As for Fig.~\ref{fig:bif-a}, except that $\delta = 0.49$: (a) $n = 3$ and (b) $n = 10$. Model~(i) remains accurate even for $n=3$, whereas Models~(ii) and~(iii) agree with the discrete chain only over limited portions of the bifurcation curve.
    }
    \label{fig:bif-b}
\end{figure}

For a cantilever rod, the critical curves associated with the first bifurcation mode, $m=1$, are compared in Figs.~\ref{fig:bif-a} and \ref{fig:bif-b}. These curves identify the combinations of the dimensionless end force $p$ and microforce $h$ at which the straight configuration bifurcates. The three homogenized models, denoted Models (i)--(iii), are compared with the original discrete chain, denoted here as Model (iv) and used as the reference system.

The four models are summarized as follows:
\begin{itemize}
\item[(i)] The equivalent rod obtained by incorporating the microforces into the homogenization process, Eq.~\eqref{eq:rotation};
\item[(ii)] The axially and flexurally homogenized rod subjected to a periodic distribution of concentrated microforces treated as external loads, Eqs.~\eqref{eq:equilibrium-post-hom-micro-force} and \eqref{eq:betaII};
\item[(iii)] The flexurally homogenized rod with piecewise-constant axial stiffness, subjected to a periodic distribution of concentrated microforces treated as external loads, Eqs.~\eqref{eq:modeliii} and \eqref{eq:newbeta};
\item[(iv)] The original discrete chain, Eq.~\eqref{eq:equilibrium-dis-detailed-cantilever-simp}.
\end{itemize}
To compare the continuous rod with discrete chains containing different numbers of cells, the total length $L$ and the effective stiffnesses $EA$ and $EI$ are kept fixed. Since the cell length is $\ell_0 = L/n$, the spring stiffnesses are scaled inversely with $\ell_0$ according to
\begin{equation}
k_a=\frac{EA}{\ell_0},
\qquad
k_r=\frac{EI}{\ell_0}.
\end{equation}
This scaling ensures that $k_a\ell_0 = EA$ and $k_r\ell_0 = EI$ for every value of $n$; consequently, the dimensionless stiffness ratio
\begin{equation}
\chi
=
\frac{k_r}{k_aL^2}
=
\frac{EI}{EA\,L^2}
\label{eq:chi}
\end{equation}
also remains constant. 

The dimensionless parameter $p$ denotes the macroscopic end load applied to the original discrete chain. The same end load is applied in Models (ii) and (iii), in which the microforces remain external loads. In Model (i), the microforces are instead included in the homogenization process, so that the corresponding effective dimensionless horizontal end force becomes $p + 2h$.

Both Figs.~\ref{fig:bif-a} and \ref{fig:bif-b} refer to $\chi = 1/25^2$. Panels (a) and (b) show the results for chains with $n = 3$ and $n = 10$ elements, respectively. The internal length ratio is $\delta = 0.3$ in Fig.~\ref{fig:bif-a} and $\delta = 0.49$ in Fig.~\ref{fig:bif-b}. 

All panels show the lines $p + 2h = -1 + 2\delta$ and $p + 2h = 0$. The first condition corresponds to the axially deformable segment of each discrete element being compressed to zero length and therefore represents a mechanical limit. The second condition corresponds to vanishing effective axial prestress and is discussed in Sec.~\ref{fifa}. 

The following conclusions can be drawn from Figs.~\ref{fig:bif-a} and \ref{fig:bif-b}.
\begin{itemize}
\item Model~(i) provides the most accurate approximation of the discrete chain, with the two responses being virtually indistinguishable for the parameter values considered. This agreement persists for very small numbers of elements, including $n = 10$ and, more remarkably, $n = 3$, even though homogenization is formally associated with the asymptotic limit $n \to \infty$.
\item Model~(ii) exhibits substantial discrepancies from the discrete chain, as expected from its failure to reproduce the correct continuum limit. Model~(iii) is generally more accurate, but agrees satisfactorily with the discrete-chain result only over a limited part of the bifurcation curve.
\item When $H=0$, or equivalently $h=0$, Model~(ii) reduces to Model~(i). Consequently, although the bifurcation curves predicted by the two models differ significantly, they intersect on the horizontal axis, where $h=0$.
\item In panels~(b), the bifurcation curve predicted by Model~(ii) appears approximately elliptical in the $h$--$p$ plane. The results in panels~(a), however, show that this shape is not a general feature of the model.
\end{itemize}

The analytical results provide further insight into the two limiting values of $\delta$. When $\delta = 0$, the microforces from adjacent cells cancel at each internal hinge, leaving only the forces applied at the two ends. The chain is therefore subjected to the effective axial load $P + 2H$. In this case, Models~(i)--(iii) reduce to the same homogeneous rod and have identical nonlinear equilibrium equations and bifurcation loads. The finite discrete chain still differs because of its discrete bending response, but its bifurcation loads converge to the common continuum prediction as $n \to \infty$.

At the other limiting value, $\delta = 1/2$, the central axial spring has zero initial length. This configuration is therefore interpreted as the limit $\delta \to 1/2^{-}$. In the discrete chain, the two microforces approach the same position but remain applied on opposite sides of the axial spring. Their application points separate when the spring elongates, so the microforces do not cancel. Model~(i) captures this effect and yields the geometric-stiffness parameter $\mathcal{F}_I^{(0)} = P+(P+2H)^2/EA$, thus reproducing the continuum limit of the chain. In Model~(ii), the microforces are instead applied at the same material point of the homogenized rod and cancel, making the predicted bifurcation loads independent of $H$. In Model~(iii), the central region shrinks to zero length while its equivalent spring stiffness remains equal to $k_a$. It therefore becomes a concentrated axial spring and retains the effect of the microforces. Since the current length of the central spring must remain positive, only the part of the bifurcation curve satisfying $P + 2H > 0$ is mechanically admissible.

These two limiting cases clarify the mechanical behavior for intermediate values of $\delta$. The agreement of Model~(ii) with the other models at $\delta = 0$ results from the cancellation of the internal microforces and does not extend to $0 < \delta < 1/2$. By contrast, Model~(iii) converges to Model~(i) as $n \to \infty$ for $0 < \delta < 1/2$, as follows from Eqs.~\eqref{eq:model-iii-continuum-condition}--\eqref{eq:G-III-G-I}, and it also recovers Model~(i) in the limit $\delta \to 1/2^{-}$. Thus, Model~(iii) reproduces the correct continuum limit over the full admissible range of $\delta$, although its predictions may differ significantly from those of the discrete chain at small $n$.

\subsection{Nonlinear behavior of the cantilever rod}
\label{seg-a}

\begin{figure}[H]
    \centering
    \includegraphics[width=0.9\linewidth]{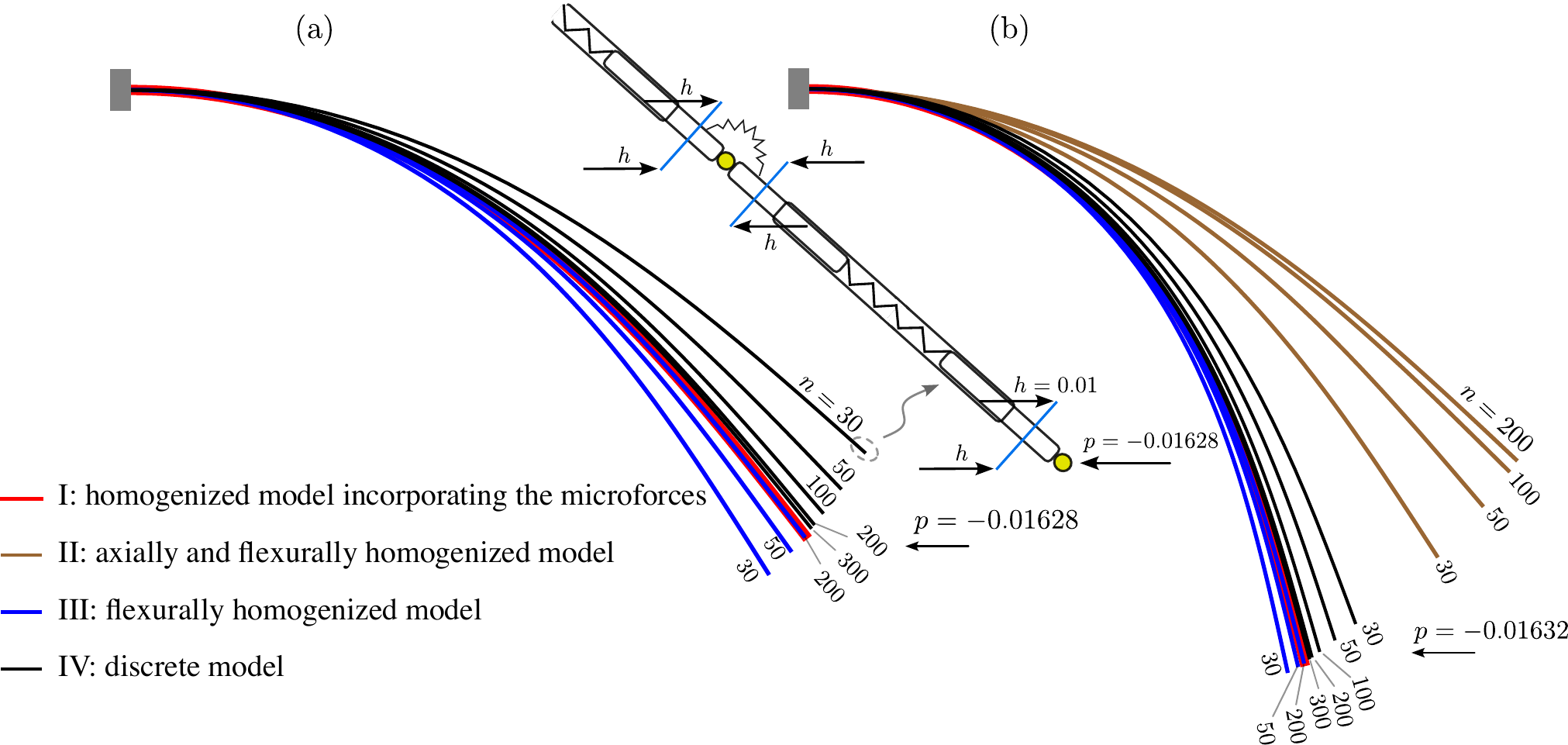}
    \caption{ Comparison of the nonlinear equilibrium configurations predicted by the rod Models~(i)--(iii) and by discrete chains with $n=30$, $50$, $100$, $200$, and $300$ elements in a cantilever configuration. The microforce parameter is $h = 0.01$, while the axial load is $p = -0.01628$ in panel~(a) and $p = -0.01632$ in panel~(b). The remaining parameters are $\chi = 10^{-4}$ and $\delta = 0.1$. Model~(i) is shown in red, Model~(ii) in brown, Model~(iii) in blue, and the discrete chains in black. Model~(ii) is not shown in panel~(a) because the applied compression has not reached its critical value for buckling, so that the rod remains straight. Model~(i) provides the closest approximation to the discrete chain. As $n$ increases, the discrete configurations converge to Model~(i) from the less-deformed side, whereas Model~(iii) converges to it from the more-deformed side. By contrast, the discrepancy associated with Model~(ii) increases with $n$. 
    }
    \label{fig:ela-comp}
\end{figure}

\begin{figure}[H]
    \centering
    \includegraphics[width=0.8\linewidth]{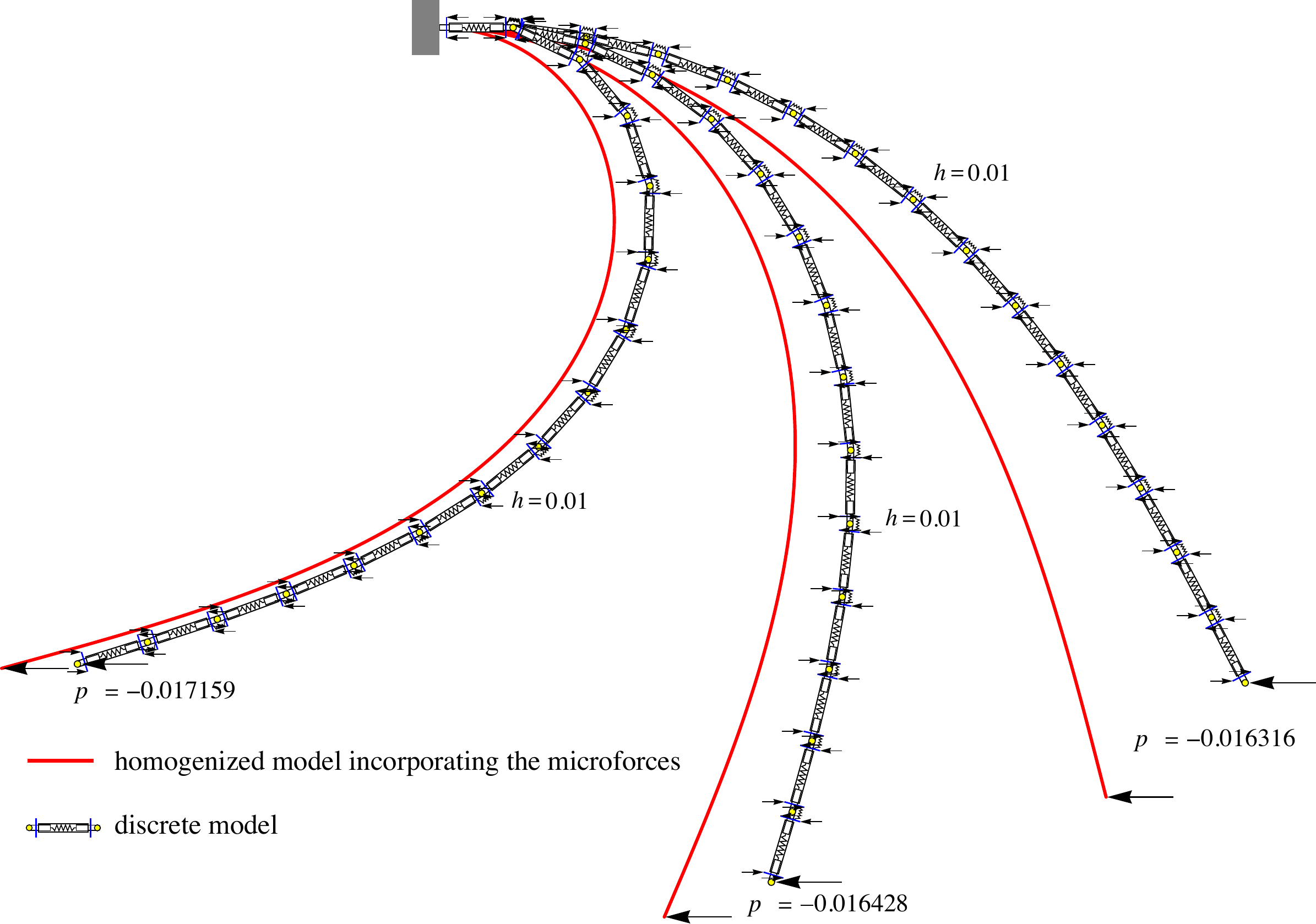}
    \caption{
    Comparison between the nonlinear equilibrium configurations predicted by the rod Model~(i), governed by Eqs.~\eqref{eq:equi-hom-sim}--\eqref{eq:axialeps}, and those of the discrete chain with $n=15$ elements, governed by the system~\eqref{eq:equilibrium-dis-detailed-cantilever-simp}. The parameters are $\chi=10^{-4}$, $\delta=0.1$, and $h=0.01$. From left to right, the dimensionless end loads are $p=-0.017159$, $p=-0.016428$, and $p=-0.016316$. The configurations predicted by Model~(i) are shown as red lines, whereas the discrete chains are shown in black, with blue rigid cross-sections and yellow rotational hinges. Each pair of configurations shares the same clamped point at the upper-left end. The arrows at the free ends and within the unit cells remain horizontal and indicate the fixed directions of the applied dead loads.
    }
\label{fig:homogenized-discrete-equilibrium-comparison}
\end{figure}

The nonlinear responses predicted by Models~(i)--(iii) are compared in Fig.~\ref{fig:ela-comp} with the corresponding response of the discrete chain, denoted as Model~(iv) and taken as the reference. A further comparison restricted to Models~(i) and~(iv) is presented in Fig.~\ref{fig:homogenized-discrete-equilibrium-comparison}. 

A cantilever subjected to the dimensionless axial dead load $p$ and the microforce parameter $h$ is considered. The squared inverse slenderness and geometric parameters are set to $\chi = 10^{-4}$ and $\delta = 0.1$. The selected load pairs in Fig.~\ref{fig:ela-comp} are $(p,h) = (-0.01628,0.01)$ and $(p,h) = (-0.01632,0.01)$ in panels (a) and (b), respectively. Figure~\ref{fig:homogenized-discrete-equilibrium-comparison} also refers to $h = 0.01$. In Fig.~\ref{fig:ela-comp}, chains with $n=30$, $50$, $100$, $200$, and $300$ elements are considered, and the corresponding value of $n$ is indicated next to each deformed configuration.

The stiffness scaling and load conventions introduced in the preceding subsection are retained in the following comparisons. Thus, $L$, $EA$, $EI$, and $\chi$ remain fixed as $n$ varies.

The post-buckling configurations were obtained as follows. For the discrete chain, the axial strains were eliminated using the axial constitutive equations and the clamp condition $\theta_1 = 0$ was imposed. The resulting $n - 1$ nonlinear rotational equilibrium equations from system~\eqref{eq:equilibrium-dis-detailed-cantilever-simp}, with $\theta_2, \ldots, \theta_n$ as unknowns, were solved using the Newton--Raphson method along the nontrivial equilibrium branch. The axial strains were then recovered from the computed rotations, and the nodal positions were reconstructed recursively using the compatibility relation~\eqref{eq:ui-slaved}. 

For Models~(ii) and~(iii), analytical solutions are not available. Their nonlinear boundary-value problems were therefore solved numerically using a shooting method implemented in Mathematica~15.0~\cite{Mathematica}. By contrast, for Model~(i), the nonlinear boundary-value problem given in dimensionless form by Eq.~\eqref{eq:equi-hom-sim-for-elliptic}, with $\xi=s/L$ and boundary conditions $\theta(0)=0$ and $\theta'(1)=0$, was solved using the closed-form solution derived in Sec.~\ref{app:solutions-elliptic}. This solution was independently verified using the same numerical shooting method.

\begin{figure}[H]
    \centering
    \includegraphics[width=0.60\linewidth]{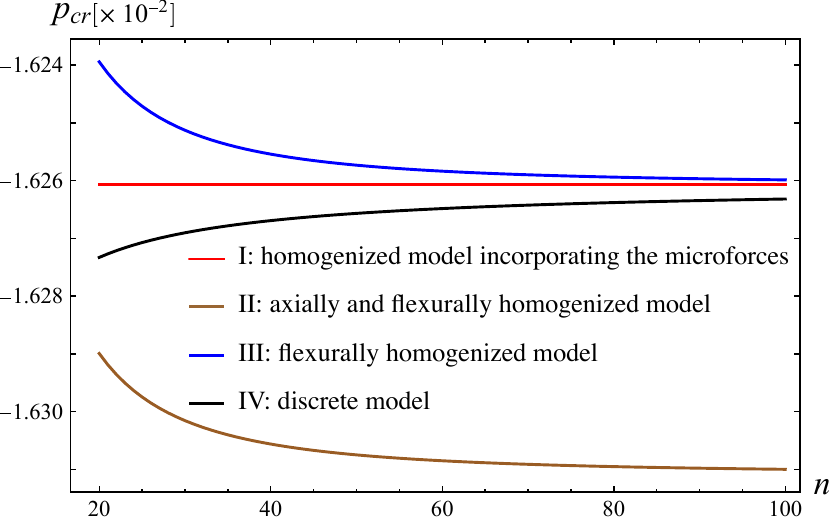}
    \caption{
    First admissible critical load for bifurcation, $p_{\mathrm{cr}}$, encountered as $p$ is decreased from zero, as a function of the number of elements $n$ for the discrete chain and the three equivalent-rod models, with $\chi=10^{-4}$, $\delta=0.1$, and $h=0.01$. The prediction of Model~(i) is independent of $n$ and gives the continuum limit approached by the discrete chain. Model~(iii) also converges to this limit from above, whereas Model~(ii) approaches a distinct, more-negative critical load.
    } 
    \label{fig:models-pcr-n-com}
\end{figure}

The comparisons in Fig.~\ref{fig:ela-comp} lead to the following observations.
\begin{itemize}
\item The deformed configuration predicted by Model~(i), governed by Eqs.~\eqref{eq:equi-hom-sim}--\eqref{eq:axialeps}, is shown in red and is independent of $n$. Because the microforces are included directly in the homogenization process, Model~(i) provides the most accurate approximation of the discrete chain. For finite $n$, the discrete chain is slightly stiffer than the equivalent rod and therefore undergoes a smaller deformation under the same loads. As $n$ increases, this difference decreases and the discrete configurations converge rapidly to the configuration predicted by Model~(i). In this sense, the convergence occurs `from above,' with the discrete configurations approaching the equivalent-rod configuration from the less-deformed side.
\item The deformed configurations predicted by Model~(ii), governed by Eqs.~\eqref{eq:equil-post-hom-micro-force}--\eqref{eq:epsilon-post-hom-micro-force}, are shown in brown in panel~(b). They are not shown in panel~(a) because the applied compression is not large enough to reach the bifurcation load predicted by Model~(ii). The rod therefore remains straight. Panel~(b) shows that the difference between Models~(ii) and~(iv) increases with $n$. As shown in Fig.~\ref{fig:models-pcr-n-com}, the magnitude of the critical compressive load predicted by Model~(ii) increases with $n$, so that a larger compression is required to produce buckling. Moreover, Model~(ii) approaches a limiting critical load different from that of the discrete chain and Model~(i). Consequently, except for the special cases in which the models coincide (when $h = 0$ or $\delta = 0$), Model~(ii) does not reproduce the correct continuum limit and provides the least accurate approximation of the discrete chain over the entire parameter range.
\item The deformed configurations predicted by Model~(iii), governed by Eqs.~\eqref{eq:equil-post-hom-micro-force-app}--\eqref{eq:epsilon-post-hom-micro-force-app}, are shown in blue. For finite $n$, Model~(iii) is more compliant than both the discrete chain and Model~(i), as confirmed by the smaller magnitude of its critical compressive load in Fig.~\ref{fig:models-pcr-n-com}. It therefore undergoes a larger deformation under the same loads. As $n$ increases, Model~(iii) becomes progressively stiffer, and its deformed configurations approach the configuration predicted by Model~(i). In this sense, the convergence occurs `from below,' with Model~(iii) approaching the Model~(i) configuration from the more-deformed side.
\end{itemize}

Figure~\ref{fig:homogenized-discrete-equilibrium-comparison} provides a further comparison between the nonlinear configurations predicted by Model~(i) and those of the discrete chain, Model~(iv), at fixed values of the loads. Only $n = 15$ discrete elements are used to demonstrate the accuracy of Model~(i), even for a relatively small number of cells. The microforce parameter is fixed at $h = 0.01$, and the three axial loads $p = -0.016316$, $-0.016428$, and $-0.017159$ are considered. These values all exceed the first critical load for buckling in compressive magnitude and therefore correspond to post-buckling equilibrium states. 

\section{Tensile buckling and restabilization}
\label{ssec:tensile-buckling-condition}

In Model~(i), the microforces are incorporated into the homogenization process, and the effective dimensionless axial resultant in the straight configuration is
\begin{equation}
p_{\mathrm{eff}}=p+2h.
\label{eq:effective-axial-resultant}
\end{equation}
Buckling is therefore classified as compressive or tensile according to the sign of $p_{\mathrm{eff}}$: compressive buckling corresponds to $p_{\mathrm{eff}}<0$, whereas tensile buckling corresponds to $p_{\mathrm{eff}}>0$. The case $p_{\mathrm{eff}}=0$, discussed in Sec.~\ref{fifa}, represents bifurcation in the absence of an effective axial resultant, so that the structure is unstretched.

For the $m$-th mode, the bifurcation condition for Model~(i) is
\begin{equation}
(p+2h)(1+p+2h)-4h\delta
+\frac{(2m-1)^2\pi^2}{4}\chi=0,
\qquad
m=1,2,\ldots ,
\label{eq:model-i-bifurcation-effective-load}
\end{equation}
an equation which, using Eq.~\eqref{eq:effective-axial-resultant}, can be rewritten as
\begin{equation}
p_{\mathrm{eff}}(1+p_{\mathrm{eff}})
=
4h\delta-\frac{(2m-1)^2\pi^2}{4}\chi.
\label{eq:effective-load-quadratic}
\end{equation}
The two solutions for the effective axial resultant at the onset of bifurcation are
\begin{equation}
p_{\mathrm{eff},m}^{\pm}
=
\frac{-1\pm
\sqrt{1+16h\delta-(2m-1)^2\pi^2\chi}}{2}.
\label{eq:effective-load-roots}
\end{equation}
A tensile bifurcation requires the positive branch in Eq.~\eqref{eq:effective-load-roots} to satisfy $p_{\mathrm{eff},m}^{+}>0$. This occurs if and only if
\begin{equation}
1+16h\delta-(2m-1)^2\pi^2\chi>1,
\end{equation}
or, equivalently,
\begin{equation}
h>
h^{\mathrm{T}}_{m}
=
\frac{(2m-1)^2\pi^2\chi}{16\delta},
\qquad
0<\delta\leq\frac{1}{2}.
\label{eq:tensile-buckling-condition}
\end{equation}
Equation~\eqref{eq:tensile-buckling-condition} is therefore the necessary and sufficient condition for the existence of a tensile bifurcation load for mode $m$. At the limiting value $h=h^{\mathrm{T}}_{m}$, the positive branch gives $p_{\mathrm{eff},m}^{+}=0$. This point corresponds to bifurcation with zero effective axial resultant and separates compressive from tensile bifurcation.

The parameter $\delta$ lies in the range $0\leq\delta\leq1/2$. When $\delta=0$, the equivalent rotational springs produced by the microforces have a vanishing stiffness. Therefore, Eq.~\eqref{eq:effective-load-quadratic} does not have a solution for $p_{\mathrm{eff}}>0$, and tensile buckling cannot occur. For $\delta>0$, the value of $h$ required for tensile buckling decreases as $\delta$ increases. The lowest threshold is obtained in the limit $\delta\to1/2^{-}$ and is
\begin{equation}
(h^{\mathrm{T}}_{m})_{\min}
=
\frac{(2m-1)^2\pi^2\chi}{8}.
\label{eq:minimum-tensile-threshold}
\end{equation}
The value $\delta=1/2$ is understood as a limiting case because the central axial spring then reaches a null initial length.

For a given value of $h$, tensile buckling is possible only for the modes satisfying
\begin{equation}
(2m-1)^2\pi^2\chi<16h\delta.
\end{equation}
This condition limits the possible values of $m$, so that only a finite number of tensile bifurcation modes can occur. The first mode, $m=1$, corresponds to the lowest threshold.

Mechanically, tensile buckling becomes possible because the homogenization of the microforces leads to a distribution of rotational springs with negative stiffness acting on the homogenized rod and this destabilizing rotational contribution, $4h\delta$, can overcome both the stabilizing effect of the tensile axial resultant and the bending stiffness. Finally, this classification refers to the effective axial resultant $p+2h$ in Model~(i). The macroscopic end load applied to the original discrete chain remains $p=p_{\mathrm{eff}}-2h$ and is negative at tensile bifurcation.

\begin{figure}[H]
    \centering
    \includegraphics[width=0.75\linewidth]{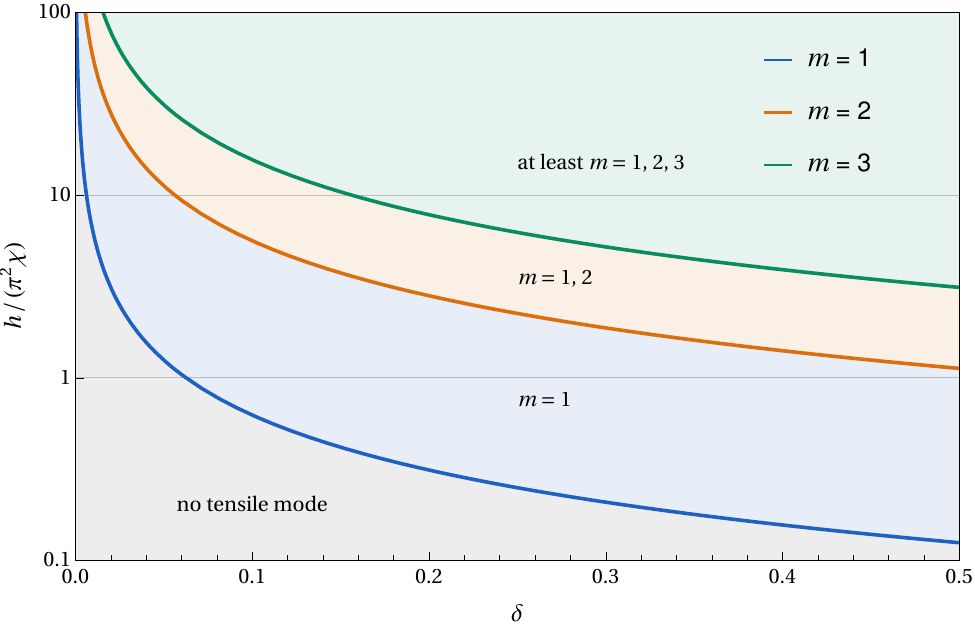}
    \caption{
    Tensile-buckling regions predicted by Model~(i) as functions of $\delta$ and the normalized microforce $h/(\pi^2\chi)$. The curves correspond to modes $m=1$, $2$, and $3$. For a given value of $\delta$, mode $m$ can undergo tensile buckling when $h/(\pi^2\chi)$ lies above the corresponding curve. The colored regions indicate which modes satisfy this condition. The vertical axis is logarithmic.
    }
    \label{fig:model_i_tensile_buckling}
\end{figure}

The tensile-buckling condition derived above is illustrated in Fig.~\ref{fig:model_i_tensile_buckling} for the first three modes. The curves divide the parameter space into regions in which tensile buckling is possible for different sets of modes.

Figure~\ref{fig:model_i_tensile_buckling} shows that for each mode, the value of $h$ required for tensile buckling decreases as $\delta$ increases. Therefore, the microforces have a stronger destabilizing effect for larger values of $\delta$. For a fixed $\delta$, the required value of $h$ increases with the mode number as $(2m-1)^2$. The first mode, therefore, requires the smallest value of $h$, and only a finite number of modes can undergo tensile buckling for a prescribed value of the magnitude of the microforces. When $\delta=0$, tensile buckling is not possible.

\begin{figure}[H]
    \centering
    \includegraphics[width=0.5\linewidth]{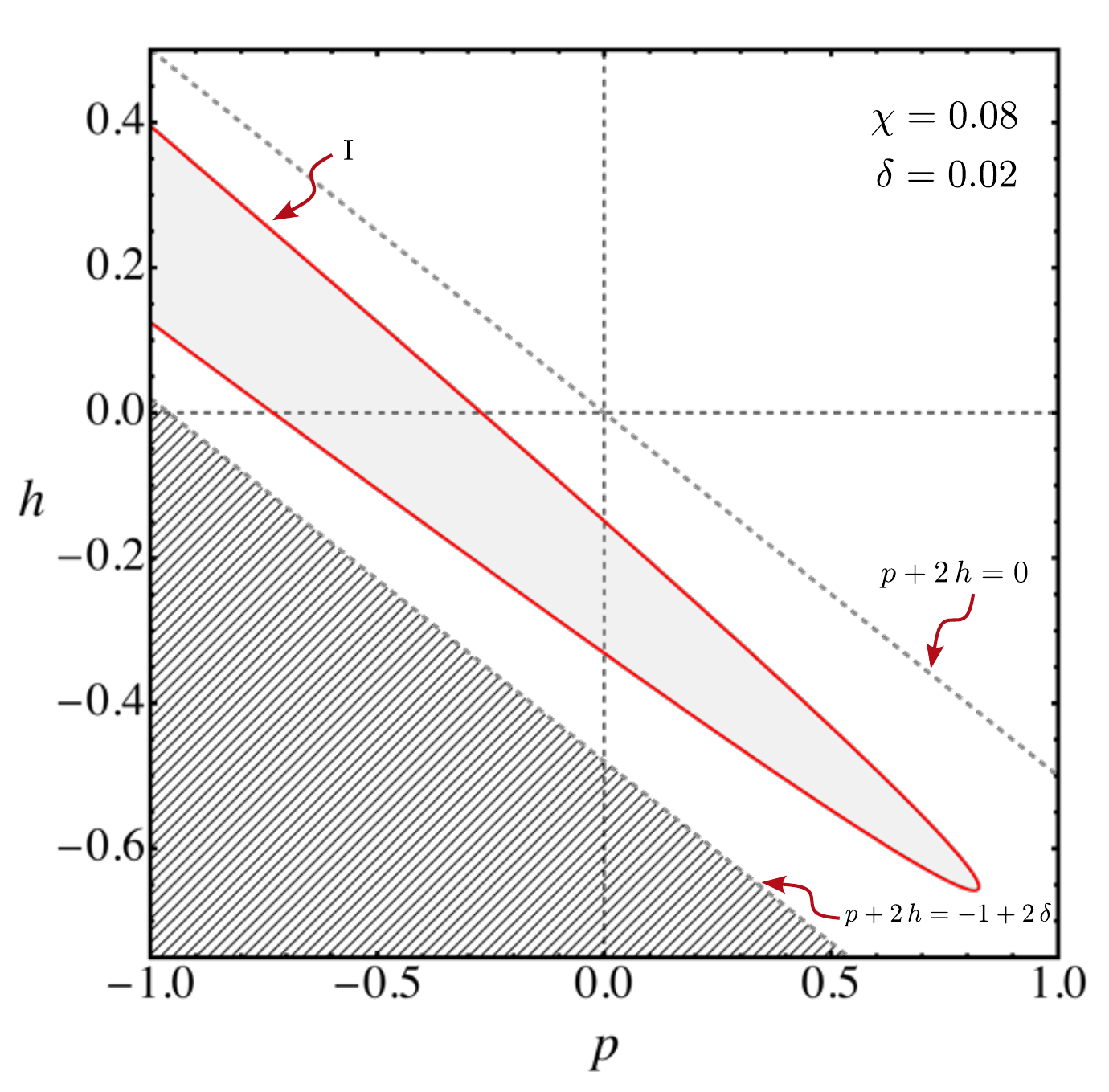}
    \caption{
    First-mode bifurcation diagram of Model~(i) in the $p$--$h$ plane for $\chi=0.08$ and $\delta=0.02$. The red curve delimits the inner gray region, where the straight configuration of the cantilever is unstable. The admissible part of the outer region is stable. The hatched region contains the load combinations that violate the mechanical condition $p+2h>-1+2\delta=-0.96$. Bifurcation curves associated with higher modes lie inside the inner gray region and are not shown. For $h=0$, the straight configuration loses stability at $p\simeq-0.2706$ and becomes stable again at $p\simeq-0.7294$, showing restabilization after compressive buckling.
    }
    \label{fig:restabilization}
\end{figure}

Besides tensile buckling, Model~(i) may also predict the recovery of stability of the straight configuration after compressive buckling, called `restabilization'. This possibility is illustrated in Fig.~\ref{fig:restabilization} for $\chi=0.08$, $\delta=0.02$, and for the first mode, $m=1$. The red curve represents the two branches of the bifurcation condition, both valid for $m=1$. They bound the inner gray region, where the straight configuration is unstable with respect to this mode. The region outside this curve is referred to as the `outer region'. Curves associated with higher modes are also present inside the inner gray region. The hatched region is mechanically inadmissible because it violates the condition $p+2h>-1+2\delta=-0.96$.

A rigorous stability analysis requires checking the sign of the second variation of the total potential energy. In the present case, however, a sufficient stability argument for the admissible outer region can be obtained from the continuity of the eigenvalues of the second variation with respect to the loading parameters. Their signs can change only when an eigenvalue vanishes, that is, on a bifurcation curve. The admissible outer region is connected, contains the unloaded configuration, and is not crossed by any bifurcation curve. Since the unloaded configuration is stable, the admissible outer region is therefore stable with respect to the considered perturbations.

The restabilization can be appreciated, for example, by setting $h=0$ and progressively decreasing $p$ from zero. The straight configuration is initially stable. It loses stability at $p\simeq-0.2706$ and remains unstable until the second bifurcation point, $p\simeq-0.7294$, where it becomes locally stable again. This second bifurcation is reached before the mechanical limit $p=-0.96$ and is therefore admissible. This behavior results from the axial deformability of the rod.

It is important to note that the second bifurcation indicates only that the straight equilibrium branch becomes locally stable again. A rod that is already following a nontrivial post-buckling branch does not necessarily return to the straight configuration at this point. Its actual response depends on the nonlinear equilibrium path and on the loading procedure.

\section{The discrete chain and its equivalent rod as a soft robot arm}
\label{sub:post-bif-cantilever-supported}

When the microforces $h$, the macroscopic end force $p$,
or both can be controlled, the proposed microstructural design enables the elastic rod to be used as a soft robotic arm. The analysis of the device is greatly facilitated by the homogenized Model (i), which provides the highest accuracy among the considered homogenization schemes while remaining sufficiently tractable to admit closed-form solutions.

In the following, \emph{load control} refers to a loading protocol in which the forces $p$ and $h$ are prescribed, whereas \emph{displacement control} refers to a protocol in which the horizontal component of the end displacement $u(1)$ is prescribed and the corresponding reaction force is allowed to adjust. The two control modes may also be combined sequentially along a loading path. When one of the loads is held fixed, its prescribed value is maintained while the other control parameter is varied.

\begin{figure}
\centering
\includegraphics[width=1\linewidth]{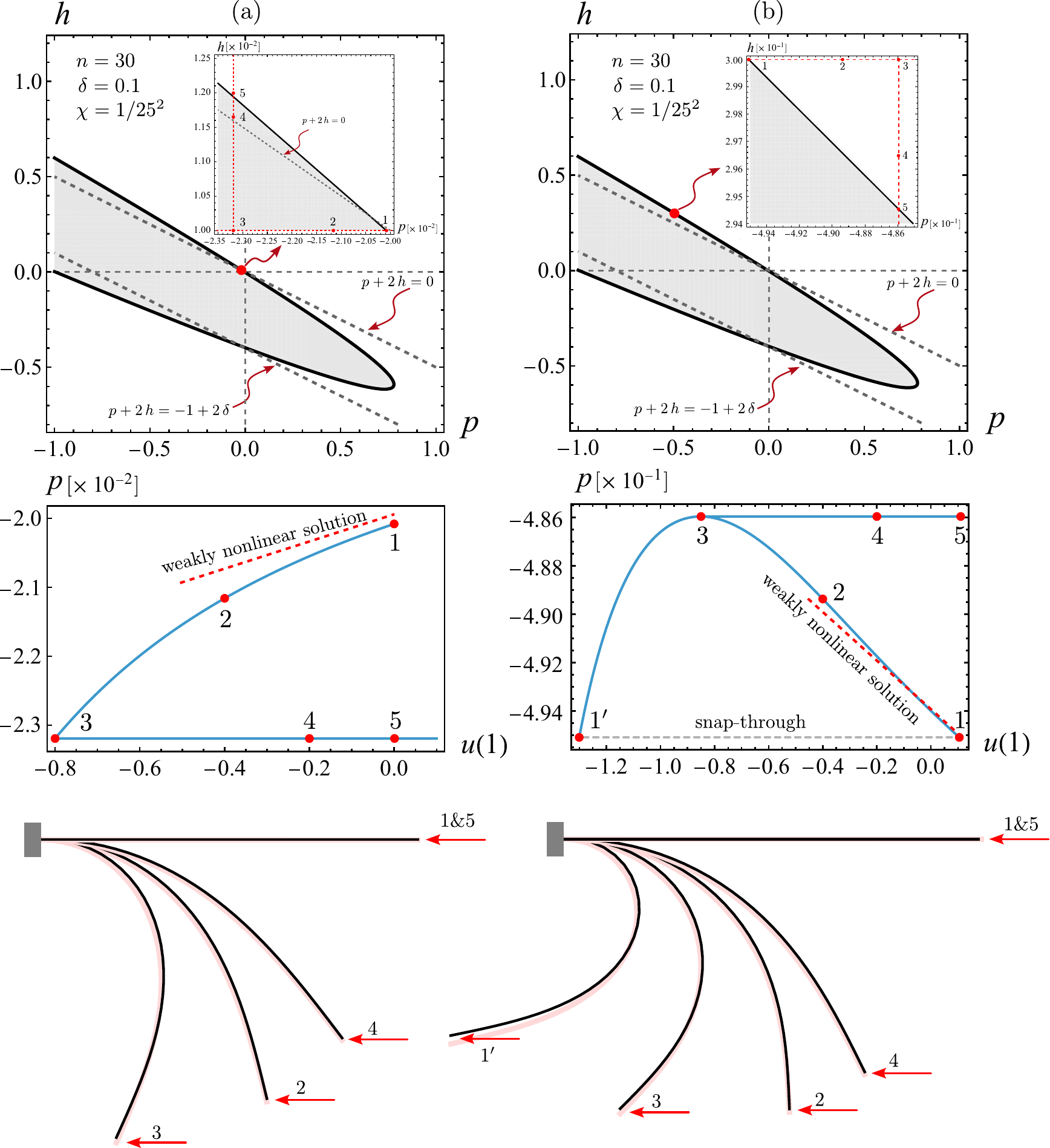}
\caption{
A microstructural chain, investigated for use as a soft robotic arm, is subjected to loading paths that connect the straight and post-buckling configurations. Top: bifurcation diagrams in the $p$--$h$ plane (gray region: the straight configuration is unstable). Middle: load $p$ versus end displacement $u(1)$, where the blue solid curves are the numerical solutions of the discrete chain and the red dashed curves are the weakly nonlinear solutions of the equivalent rod. Bottom: deformed configurations of the discrete chain (black) and equivalent continuous model (light red). In (a), increasing $|p|$ followed by increasing $|h|$ produces a supercritical bifurcation, allowing the entire deformation cycle to be achieved under load control. In (b), the weakly nonlinear solution (red dashed curve) predicts a subcritical bifurcation at state (1). Under load control, the system therefore jumps from state (1) to the stable state (1'), whereas displacement control is required to follow the post-critical branch continuously through states (2) and (3). From state (3), load control can be resumed by holding $p$ fixed while decreasing $h$, leading through states (4) and (5) toward the straight configuration.
}
\label{fig:combined_supercritical_subcritical_postbuckling}
\end{figure}

To illustrate the potential of the structure for soft-robotic applications, two representative loading paths are considered in Fig.~\ref{fig:combined_supercritical_subcritical_postbuckling}. The results are obtained by comparing the numerical solutions of the discrete chain (black curve), with $n=30$ elements, with the analytical solution of the equivalent rod (red-light curve), characterized by $\chi=1/25^2$ and $\delta=0.1$. The loading paths shown in the upper panels correspond to portions of the stability regions in the $p$--$h$ plane, similar to those reported in Figs.~\ref{fig:bif-a} and \ref{fig:bif-b}. The black curve denotes the bifurcation boundary given by Eq.~\eqref{eq:p-critical-discrete} for $m=1$, while the dashed lines correspond, respectively, to the vanishing of the axial prestress, $p+2h=0$, and to the compressive annihilation of the springs in the chain, $p+2h+1=2\delta$.

The post-buckling configurations of the equivalent rod shown in the lower panels are obtained from the analytical solution reported in Sec.~\ref{app:solutions-elliptic}. A weakly nonlinear solution is also developed in Appendix~\ref{app:weakly-nonlinear-clamped-beam} to verify the analytical results and to characterize the initial post-critical behavior of the homogenized structure.
The analysis provides only local information near the bifurcation point and suggests possible equilibrium-path scenarios, without establishing their stability or the precise endpoint of snap-through, both of which may depend on imperfections and dynamic effects.

In the left-hand panels of Fig.~\ref{fig:combined_supercritical_subcritical_postbuckling}, the compressive macroscopic end force $p$ is increased in magnitude while the microforce is kept fixed at $h=0.01$. The system, therefore, evolves under load control from the initial straight configuration through the bifurcation and into the post-buckling regime, corresponding to states (1)--(3). Subsequently, $p$ is held fixed while the magnitude of $h$ is increased, again under load control, bringing the system back to the straight configuration through states (4) and (5). The weakly nonlinear analysis shows that the bifurcation at state (1) is supercritical, so that the stable post-buckling branch emerges continuously from the bifurcation point. Consequently, the entire deformation cycle illustrated in panel (a) can be achieved by controlling the applied loads $p$ and $h$.

In the right-hand panels of Fig.~\ref{fig:combined_supercritical_subcritical_postbuckling}, the initial loading is likewise performed under load control, with $h$ held fixed and $|p|$ increased. In this case, however, the bifurcation is subcritical. The post-critical branch emerging from the bifurcation point is therefore unstable, and the system cannot follow this branch continuously under load control. When the critical load is reached, the end displacement $u(1)$ consequently jumps from the initial state (1) to the stable post-buckling state (1'). To access the intermediate post-critical states (2) and (3), the control mode must instead be switched to displacement control, with $h$ held fixed and $u(1)$ prescribed. The system can then be driven continuously along the post-critical branch. Once state (3) has been reached, load control can be resumed by holding $p$ fixed and decreasing $h$, leading through states (4) and (5) back toward the straight configuration.

The occurrence of a subcritical bifurcation in the cantilevered elastic rod is unexpected and results from the combined action of the macroscopic end load $p$ and the microforces $h$. The stable or unstable character of the bifurcation point is determined by the sign of the coefficient $c_3^{\mathrm{cr}}$ in Eq.~\eqref{eq:bifurcation-classified}, derived in Appendix~\ref{app:weakly-nonlinear-clamped-beam}. Thus, controlling the microforces in addition to the axial load provides a means of modifying the post-critical response and, in particular, of changing the character of the bifurcation from supercritical to subcritical, or vice versa. This feature provides an additional mechanism for controlling the deformation and stability of the proposed soft robotic arm.

\section{Conclusions}
\label{sec:concluding}

Inspired by applications in soft robotics, a microstructured elastic chain has been introduced in which, in addition to end forces, a longitudinal distribution of dead microforces is applied. These microforces have a zero resultant in the straight configuration, where the chain is subjected to a piecewise-constant axial force distribution.

The discrete model has been homogenized using three different schemes, leading to a result of particular theoretical interest. The most effective homogenization strategy yields an equivalent elastic rod in which the microforces no longer appear explicitly as external loads. Instead, their effect is incorporated into an internal prestress, determined by the boundary conditions, together with a distributed rotational-spring contribution acting along the rod.
The resulting homogenized model, therefore, provides mechanical conditions that are not readily represented within conventional rod formulations. In particular, suitable combinations of the applied loads can generate effective rotational springs with negative stiffness.

Beyond providing a new homogenization strategy for a class of microstructured systems, the formulation makes available analytical tools that cannot be applied directly to the discrete model or to alternative homogenized formulations in which the microforces remain explicit external loads. In particular, the equivalent rod admits closed-form solutions for large deflections, well within the nonlinear regime, and provides a convenient framework for weakly nonlinear analysis of the initial post-critical response.

The results also demonstrate the potential of the proposed structure for soft-robotic applications. The homogenized rod exhibits both tensile and compressive buckling, with restabilization occurring in the compressive regime. Moreover, suitable loading paths allow deformation cycles in which the straight configuration is first abandoned and subsequently recovered. The combination of axial loading and controllable microforces also enables modification of the bifurcation character, leading to either supercritical or subcritical post-buckling behavior. These features highlight the potential of the proposed microstructured rod as a mechanically programmable soft robotic arm.

%

\section*{Acknowledgements}

D.B. and Y.Y. acknowledge funding from the European Research Council (ERC) under the European Union’s Horizon Europe research and innovation programme, Grant agreement No. ERC-ADG-2021-101052956-BEYOND. 
A.P. acknowledges funding from the European Research Council (ERC) under the European Union’s Horizon Europe research and innovation programme, Grant agreement No. HE~GA~101086644~S-FOAM. 
Views and opinions expressed are those of the authors only and do not necessarily reflect those of the European Union or the European Research Council Executive Agency. Neither the European Union nor the granting authority can be held responsible for them. 
The methodologies developed in this work fall within the aims of the GNFM (Gruppo Nazionale per la Fisica Matematica) of the INDAM (Istituto Nazionale di Alta Matematica).

\appendix

\section{Transfer Matrix}
\label{app:Transfer-Matrix}

The transfer matrices used in Sec.~\ref{zorro} to determine the buckling load of a straight elastic rod subjected to an axial load and a discrete distribution of microforces along its axis are derived below.

\subsection{Transfer Matrix for a unit cell}
\label{sub_app:Transfer-Matrix-unit-cell}

Consider a segment of an elastic rod of length $\ell$, with prescribed rotation and curvature at the left endpoint $s=0$, collected in the state vector $\bm{y}(0)$ as
\begin{equation}
\bm{y}(0) = \{ \theta(0), \theta'(0) \}^{\mathsf T}.
\end{equation}
The linearized equilibrium equation for the rod is
\begin{equation}
\theta''(s) + \beta^2 \theta(s) = 0 ,
\end{equation}
whose solution is
\begin{equation}
\theta(s) = \theta(0) \cos(\beta s) + \frac{\theta'(0)}{\beta} \sin(\beta s). 
\end{equation}
The state vector $\bm{y}(\ell)= \{\theta(\ell), \theta'(\ell)\}^{\mathsf T}$ at $s = \ell$ then reads as 
\begin{equation}
\bm{y}(\ell)
= 
\bm{\Gamma}(\beta,\ell)\,\bm{y}(0),
\end{equation}
where
\begin{equation}
\bm{\Gamma}(\beta,\ell)
=
\begin{pmatrix}
\cos(\beta\,\ell) & \dfrac{\sin(\beta\,\ell)}{\beta} \\[3mm]
-\beta\,\sin(\beta\,\ell) & \cos(\beta\,\ell)
\end{pmatrix} 
\end{equation}
is the transfer matrix, Eq.~(\ref{eq:Transfer-Matrix}). The transfer matrix has a unit determinant, $\det \bm{\Gamma} = 1$.

\subsection{Transfer Matrix for \texorpdfstring{$n$}{n} unit cells}
\label{sub_app:Transfer-Matrix-n-unit-cell}

Note that the transfer matrix $\widetilde{\bm{\Gamma}}$, Eq.~\eqref{eq:tildegamma}, satisfies 
\begin{equation}
| \widetilde{\bm{\Gamma}} | = \tilde{\Gamma}_{11}  \tilde{\Gamma}_{22}
- \tilde{\Gamma}_{21}  \tilde{\Gamma}_{12} = 1. 
\end{equation}
Its characteristic polynomial is
\begin{equation}
\left| \widetilde{\bm{\Gamma}} - \Lambda \bm{I}_2 \right| = \Lambda^2 - 2 \tilde{\Gamma}_{22} \Lambda + 1,
\label{Ttilda-eigenvalues}
\end{equation}
so that $\widetilde{\bm{\Gamma}}$ satisfies the Cayley-Hamilton theorem 
\begin{equation}
\widetilde{\bm{\Gamma}}^{2}
- 2\, \tilde{\Gamma}_{22}\, \widetilde{\bm{\Gamma}}
+ \bm{I}_2
= \bm{0}.
\label{eq:T-tilda-eq}
\end{equation}
Multiplying Eq.~\eqref{eq:T-tilda-eq} by $\widetilde{\bm{\Gamma}}^{\,n-2}$ yields
\begin{equation}
\widetilde{\bm{\Gamma}}^{\,n}
- 2\, \tilde{\Gamma}_{22}\, \widetilde{\bm{\Gamma}}^{\,n-1}
+ \widetilde{\bm{\Gamma}}^{\,n-2}
= \bm{0}.
\label{eq:T-tilda-iteration}
\end{equation}
Since $\widetilde{\bm{\Gamma}}^{\,0} = \bm{I}_2$ (the 2$\times$2 identity matrix) and $\widetilde{\bm{\Gamma}}^{\,1} = \widetilde{\bm{\Gamma}}$, a repeated application of Eq.~\eqref{eq:T-tilda-iteration} shows that $\widetilde{\bm{\Gamma}}^{\,n}$ can be expressed as a linear combination of $\bm{I}_2$ and $\widetilde{\bm{\Gamma}}$ as
\begin{equation}
\widetilde{\bm{\Gamma}}^{\,n}
=
\mathcal{C}_n \widetilde{\bm{\Gamma}}
+
\mathcal{D}_n \bm{I}_2 ,
\label{eq:T-tilda-decompisite}
\end{equation}
where the coefficients depend only on $n$. Substituting Eq.~\eqref{eq:T-tilda-decompisite} into Eq.~\eqref{eq:T-tilda-iteration} gives
\begin{equation}
\left( \mathcal{C}_n \widetilde{\bm{\Gamma}} + \mathcal{D}_n \bm{I}_2 \right)
- 2\, \tilde{\Gamma}_{22}\,
\left( \mathcal{C}_{n-1} \widetilde{\bm{\Gamma}} + \mathcal{D}_{n-1} \bm{I}_2 \right)
+ \left( \mathcal{C}_{n-2} \widetilde{\bm{\Gamma}} + \mathcal{D}_{n-2} \bm{I}_2 \right)
= \bm{0}.
\end{equation}
Equating the coefficients of $\widetilde{\bm{\Gamma}}$ and $\bm{I}_2$ leads to
\begin{equation}
\mathcal{C}_n  =  2\, \tilde{\Gamma}_{22}\, \mathcal{C}_{n-1} - \mathcal{C}_{n-2},
\qquad
\mathcal{D}_n  =  2\, \tilde{\Gamma}_{22}\, \mathcal{D}_{n-1} - \mathcal{D}_{n-2},
\label{eq:Cn-Dn}
\end{equation}
subject to the boundary conditions
\begin{equation}
\{\mathcal{C}_0, \mathcal{C}_1\} = \{0,1\}, \hspace{3em} \{\mathcal{D}_0, \mathcal{D}_1\} = \{1,0\}.
\end{equation}
Equation~\eqref{eq:Cn-Dn} defines the recurrence relation of the Chebyshev polynomials of the second kind, so that the representation follows
\begin{equation}
\mathcal{C}_n = \frac{\sin(n\mu)}{\sin\mu}, \hspace{5em} \mathcal{D}_n = -\frac{\sin((n-1)\mu)}{\sin\mu}.
\label{eq:sol-Cn-Dn}
\end{equation}
Substituting Eq.~\eqref{eq:sol-Cn-Dn} into $\widetilde{\bm{\Gamma}}^n$ in Eq.~\eqref{eq:T-tilda-decompisite} yields
\begin{equation}
\widetilde{\bm{\Gamma}}^n  =  \dfrac{\sin (n \mu)}{\sin \mu} 
\widetilde{\bm{\Gamma}} - \dfrac{\sin ((n - 1) \mu)}{\sin \mu} 
\bm{I}_2 ,
\end{equation}
which can be rearranged to give Eq.~(\ref{mukka}).

\section{Weakly nonlinear analysis of the equivalent cantilever rod}
\label{app:weakly-nonlinear-clamped-beam}

The weakly nonlinear analysis used to obtain the results reported in Fig.~\ref{fig:combined_supercritical_subcritical_postbuckling} is developed below for the equivalent rod governed by Eq.~\eqref{eq:equi-hom-sim}, in a cantilever configuration. In Model~(i), the macroscopic end-load parameter $p$ and the microforce parameter $h$ produce the effective horizontal end load $p+2h$. The analysis is restricted to the path $h=h_{\mathrm{cr}}$, along which $h$ is fixed at its critical value and $p$ is varied near $p_{\mathrm{cr}}$. The analysis first determines whether the bifurcation is supercritical or subcritical. The corresponding weakly nonlinear relation between $p$ and the horizontal free-end displacement $u(1)$ is then derived and used to plot the red dashed curves in Fig.~\ref{fig:combined_supercritical_subcritical_postbuckling}.

A Taylor expansion about the straight configuration, $\theta=0$, of the dimensionless equilibrium equation~\eqref{eq:equi-hom-sim-for-elliptic} gives
\begin{equation}
\theta''(\xi)
-\frac{1}{\chi}
\left\{
(p+2h)+(p+2h)^2
\left[1-\frac{\theta^2(\xi)}{2}+\mathcal{O}(\theta^4)\right]
-4h\delta
\right\}
\left[\theta(\xi)-\frac{\theta^3(\xi)}{6}+\mathcal{O}(\theta^5)\right]
=0,
\label{eq:hom-equil-cantilever-weakly-a}
\end{equation}
which can be written as
\begin{equation}
\theta''(\xi)+c_1\,\theta(\xi)+c_3\,\theta^3(\xi)
+\mathcal{O}(\theta^5)=0,
\label{eq:hom-equil-cantilever-weakly-b}
\end{equation}
where
\begin{equation}
c_1(p,h)
=-\frac{1}{\chi}
\left[(p+2h)(1+p+2h)-4h\delta\right],
\qquad
c_3(p,h)
=\frac{(p+2h)^2}{2\chi}-\frac{c_1(p,h)}{6}.
\end{equation}
Let $(p_{\mathrm{cr}},h_{\mathrm{cr}})$ denote a critical point associated with the first cantilever mode and define
$c_1^{\mathrm{cr}}=c_1(p_{\mathrm{cr}},h_{\mathrm{cr}})$ and
$c_3^{\mathrm{cr}}=c_3(p_{\mathrm{cr}},h_{\mathrm{cr}})$. Equation~\eqref{eq:hom-equil-cantilever-weakly-b} can then be rewritten as
\begin{equation}
\mathcal{L}\theta
+(c_1-c_1^{\mathrm{cr}})\theta
+c_3\theta^3
+\mathcal{O}(\theta^5)=0,
\qquad
\mathcal{L}=\frac{d^2}{d\xi^2}+c_1^{\mathrm{cr}}.
\label{eq:hom-equil-cantilever-weakly-balance}
\end{equation}
Along the fundamental equilibrium path, the rotation is identically zero. Assuming that a nontrivial solution $\theta(\xi;p)$ intersects the fundamental solution at $p=p_{\mathrm{cr}}$, the direction of the buckling mode is defined, following Budiansky~\cite{Budiansky1974}, by
\begin{equation}
\frac{\phi_{\mathrm{cr}}}{\lVert\phi_{\mathrm{cr}}\rVert}
=\lim_{p\to p_{\mathrm{cr}}}
\frac{\theta}{\lVert\theta\rVert},
\label{eq:buckling-mode-limit}
\end{equation}
where the limit is taken along the bifurcated path and $\lVert\cdot\rVert$ denotes a suitable norm. The first buckling mode is unique up to its sign. The sign is chosen so that $\phi_{\mathrm{cr}}(1)>0$, and the mode is normalized by $\phi_{\mathrm{cr}}(1)=1$.

Dividing the equilibrium equation~\eqref{eq:hom-equil-cantilever-weakly-balance} by $\lVert\theta\rVert$ and letting $p\to p_{\mathrm{cr}}$ gives the bifurcation problem
\begin{equation}
\mathcal{L}\phi_{\mathrm{cr}}=0.
\label{eq:critical-mode-problem}
\end{equation}
For the boundary conditions $\phi_{\mathrm{cr}}(0)=0$ and $\phi_{\mathrm{cr}}'(1)=0$, its first solution is
\begin{equation}
\phi_{\mathrm{cr}}(\xi)
=\sin\left(\frac{\pi\xi}{2}\right),
\qquad
c_1^{\mathrm{cr}}
=\left(\frac{\pi}{2}\right)^2.
\label{eq:first-critical-mode}
\end{equation}
At the critical point, the operator $\mathcal{L}$, equipped with the homogeneous boundary conditions, is singular, with
\begin{equation}
\ker\mathcal{L}
=\operatorname{span}\{\phi_{\mathrm{cr}}\}.
\end{equation}
This singularity leads to the solvability condition used at order $\mathcal{O}(a^3)$.

Following Budiansky~\cite{Budiansky1974}, the scalar parameter $a$ is now introduced as
\begin{equation}
a=
\frac{\displaystyle\int_0^1
\theta(\xi)\phi_{\mathrm{cr}}(\xi)\,d\xi}
{\displaystyle\int_0^1
\phi_{\mathrm{cr}}^2(\xi)\,d\xi}.
\label{eq:signed-modal-amplitude}
\end{equation}
The parameter $a$ therefore measures the amount of the buckling mode contained in the rotation $\theta$. It is a signed quantity: the equilibrium equation and the boundary conditions are invariant under $\theta\mapsto-\theta$, which corresponds to $a\mapsto-a$. Thus, the system admits two symmetric post-buckling solutions which have opposite rotations but the same value of $p$. Consequently, $\theta$ is an odd function of $a$, whereas $p$ is an even function of $a$. For $|a|\ll1$, the expansions take the form
\begin{equation}
\begin{aligned}
\theta(\xi)
&=a\phi_{\mathrm{cr}}(\xi)
+a^3\theta_3(\xi)+\mathcal{O}(a^5),\\[2mm]
p
&=p_{\mathrm{cr}}+a^2p_2
+\mathcal{O}(a^4),\\[2mm]
h&=h_{\mathrm{cr}},
\end{aligned}
\label{eq:weakly-nonlinear-expansion}
\end{equation}
where $p_2$ is the initial post-buckling coefficient. The third-order correction $\theta_3$ satisfies the homogeneous boundary conditions $\theta_3(0)=0$ and $\theta_3'(1)=0$, together with
$\int_0^1\theta_3\phi_{\mathrm{cr}}\,d\xi=0$. With the adopted normalization,
$\theta(1)=a+\mathcal{O}(a^3)$, and the two symmetric bifurcated configurations correspond to $a>0$ and $a<0$.

\paragraph{Critical point and load expansion.}
For the first cantilever mode, $c_1^{\mathrm{cr}}=\pi^2/4$. Along the path $h=h_{\mathrm{cr}}$, $p$ is decreased from a stable state lying above both critical roots, so that the first critical point corresponds to the larger root of the bifurcation equation, namely
\begin{equation}
p_{\mathrm{cr}}+2h_{\mathrm{cr}}
=-\frac{1}{2}
+\frac{1}{2}
\sqrt{1-\pi^2\chi+16h_{\mathrm{cr}}\delta}
\geq-\frac{1}{2},
\end{equation}
where the equality holds only when the two roots coincide. For convenience, define
\begin{equation}
S=1+2(p_{\mathrm{cr}}+2h_{\mathrm{cr}})
=\sqrt{1-\pi^2\chi+16h_{\mathrm{cr}}\delta}\geq0.
\end{equation}
The analysis below applies to a simple bifurcation point, i.e. $S>0$.
Substitution of the expansion of $p$ in Eq.~\eqref{eq:weakly-nonlinear-expansion} into $c_1(p,h_{\mathrm{cr}})$ and $c_3(p,h_{\mathrm{cr}})$ gives
\begin{equation}
c_1
=c_1^{\mathrm{cr}}
-\frac{S}{\chi}p_2a^2
+\mathcal{O}(a^4),
\qquad
c_3
=c_3^{\mathrm{cr}}+\mathcal{O}(a^2).
\end{equation}
If the two critical roots coincide, then $S=0$ and the linear variation of $c_1$ with $p$ vanishes. In this case, the scaling of $p-p_{\mathrm{cr}}$ must be reconsidered.

\paragraph{Order $\mathcal{O}(a^3)$.}
At cubic order, the equilibrium equation becomes
\begin{equation}
\mathcal{L}\theta_3
=\frac{S}{\chi}p_2\phi_{\mathrm{cr}}
-c_3^{\mathrm{cr}}\phi_{\mathrm{cr}}^3.
\label{eq:hom-equil-cantilever-weakly-first}
\end{equation}
Since $\mathcal{L}$ is singular and self-adjoint, the Fredholm solvability condition is
\begin{equation}
\int_0^1
\phi_{\mathrm{cr}}
\left(
\frac{S}{\chi}p_2\phi_{\mathrm{cr}}
-c_3^{\mathrm{cr}}\phi_{\mathrm{cr}}^3
\right)d\xi=0.
\end{equation}
Using
$\int_0^1\phi_{\mathrm{cr}}^2\,d\xi=1/2$ and
$\int_0^1\phi_{\mathrm{cr}}^4\,d\xi=3/8$ gives
\begin{equation}
p_2=\frac{3\chi c_3^{\mathrm{cr}}}{4S}.
\label{eq:hom-equil-cantilever-weakly-first-sol}
\end{equation}
Therefore,
\begin{equation}
p-p_{\mathrm{cr}}
=\frac{3\chi c_3^{\mathrm{cr}}}{4S}\,a^2
+\mathcal{O}(a^4).
\label{eq:p-amplitude-weakly-nonlinear}
\end{equation}
For $c_3^{\mathrm{cr}}\neq0$, the amplitude satisfies
\begin{equation}
a^2
=\frac{4S}{3\chi c_3^{\mathrm{cr}}}
\left(p-p_{\mathrm{cr}}\right)
+\mathcal{O}\!\left[
\left(p-p_{\mathrm{cr}}\right)^2
\right].
\label{eq:amplitudeSquare}
\end{equation}
\paragraph{Bifurcation type.}
The coefficient $p_2$ plays the role of the initial post-buckling coefficient in the symmetric-bifurcation expansion. Since $S>0$ and $a^2>0$ on a nontrivial branch, its sign is the same as that of $c_3^{\mathrm{cr}}$. In the present convention, increasing compression corresponds to decreasing $p$. The bifurcation is therefore classified as
\begin{equation}
\left\{
\begin{array}{ll}
c_3^{\mathrm{cr}}<0,
&\text{supercritical bifurcation, with }p<p_{\mathrm{cr}},\\[1mm]
c_3^{\mathrm{cr}}=0,
&\text{degenerate threshold; fifth-order terms are required},\\[1mm]
c_3^{\mathrm{cr}}>0,
&\text{subcritical bifurcation, with }p>p_{\mathrm{cr}}.
\end{array}
\right.
\label{eq:bifurcation-classified}
\end{equation}
This classification concerns the first loss-of-stability point. At the second critical point associated with restabilization, the smaller root satisfies $p_{\mathrm{cr}}+2h_{\mathrm{cr}}<-1/2$, and the direction of the branch with respect to $p$ is reversed.

\paragraph{Weakly nonlinear load--displacement relation.}
To represent the initial nontrivial branch in the $p$--$u(1)$ plane, the modal amplitude is now related to the horizontal displacement of the free end. Integrating Eq.~\eqref{eq:uprime-vprime}$_1$, the axial displacement is given by
\begin{equation}
u(\xi)
=\int_0^\xi
\left\{
[1+\varepsilon(\zeta)]\cos\theta(\zeta)-1
\right\}d\zeta,
\end{equation}
because $u(0)=0$. Since $h=h_{\mathrm{cr}}$ and the axial strain is $\varepsilon=(p+2h_{\mathrm{cr}})\cos\theta$, the displacement of the free end is
\begin{equation}
\begin{aligned}
u(1)
&=\int_0^1
\left\{
[1+(p+2h_{\mathrm{cr}})\cos\theta(\xi)]\cos\theta(\xi)-1
\right\}d\xi\\[1mm]
&=\int_0^1
\left\{
p+2h_{\mathrm{cr}}
-\left(p+2h_{\mathrm{cr}}+\frac{1}{2}\right)\theta^2(\xi)
+\mathcal{O}(\theta^4)
\right\}d\xi\\[1mm]
&=p+2h_{\mathrm{cr}}
-\frac{1}{2}
\left(p_{\mathrm{cr}}+2h_{\mathrm{cr}}+\frac{1}{2}\right)a^2
+\mathcal{O}(a^4).
\end{aligned}
\label{eq:u1}
\end{equation}
In the coefficient of $a^2$, $p$ can be evaluated at $p_{\mathrm{cr}}$ because $p-p_{\mathrm{cr}}=\mathcal{O}(a^2)$. The leading term, however, must retain the current value $p+2h_{\mathrm{cr}}$. This term is the displacement of the straight branch at the current load and accounts for the prebuckling axial deformation.

Let $\tilde{u}=u(1)$. At the critical point, the nontrivial branch meets the straight branch and $a=0$. Therefore, Eq.~\eqref{eq:u1} gives
\begin{equation}
\tilde{u}_{\mathrm{cr}}
\equiv\tilde{u}(p_{\mathrm{cr}},h_{\mathrm{cr}})
=p_{\mathrm{cr}}+2h_{\mathrm{cr}}.
\end{equation}
Along the path $h=h_{\mathrm{cr}}$, Eq.~\eqref{eq:u1} becomes
\begin{equation}
\tilde{u}-\tilde{u}_{\mathrm{cr}}
=(p-p_{\mathrm{cr}})-\frac{S}{4}a^2
+\mathcal{O}(a^4).
\end{equation}
For $c_3^{\mathrm{cr}}\neq0$, eliminating $a^2$ using Eq.~\eqref{eq:p-amplitude-weakly-nonlinear} gives
\begin{equation}
\tilde{u}-\tilde{u}_{\mathrm{cr}}
=\left(
1-\frac{S^2}{3\chi c_3^{\mathrm{cr}}}
\right)
(p-p_{\mathrm{cr}})
+\mathcal{O}\!\left[(p-p_{\mathrm{cr}})^2\right].
\end{equation}
Equivalently, provided that
$3\chi c_3^{\mathrm{cr}}\neq S^2$,
\begin{equation}
\begin{aligned}
p-p_{\mathrm{cr}}
={}&g(p_{\mathrm{cr}},h_{\mathrm{cr}})
\left(\tilde{u}-\tilde{u}_{\mathrm{cr}}\right)
+\mathcal{O}\!\left[
\left(\tilde{u}-\tilde{u}_{\mathrm{cr}}\right)^2
\right],\\[1mm]
g(p_{\mathrm{cr}},h_{\mathrm{cr}})
={}&
\frac{3\chi c_3^{\mathrm{cr}}}
{3\chi c_3^{\mathrm{cr}}-S^2}.
\end{aligned}
\label{eq:p-u-weakly-nonlinear-sol}
\end{equation}
Equation~\eqref{eq:p-u-weakly-nonlinear-sol} provides the local weakly nonlinear approximation of the initial nontrivial equilibrium path in the $p$--$u(1)$ plane. This relation is used to plot the red dashed curves in Fig.~\ref{fig:combined_supercritical_subcritical_postbuckling}.

\printbibliography

\end{document}